\documentclass[aps,prb,groupedaddress,floatfix,twocolumn]{revtex4}

\usepackage{graphicx}
\usepackage{amsbsy,amssymb}
\usepackage{amsmath}
\usepackage{epsf}
\usepackage[usenames,dvips]{color}
\newcommand{\be}{\begin{equation}}
\newcommand{\ee}{\end{equation}}
\newcommand{\bea}{\begin{eqnarray}}
\newcommand{\eea}{\end{eqnarray}}
\newcommand{\bw}{\begin{widetext}}
\newcommand{\ew}{\end{widetext}}
\newcommand{\LCO}{La$_2$CuO$_4$}
\newcommand{\CuO}{CuO$_2$}

\newcommand{\bk}{\boldsymbol {k}}
\newcommand{\bq}{\boldsymbol {q}}
\newcommand{\fl}{\hspace*{-.1cm}}

\begin{document}

%\preprint{QCMP Theory, 03-06-1}

\title{Magnetic susceptibility
        of a CuO$_2$ plane in the La$_2$CuO$_4$ system:\\
        I. RPA treatment of the Dzyaloshinskii-Moriya Interactions}



\author{K. V. Tabunshchyk}
\email{tkir@physics.queensu.ca}
\affiliation{Department of Physics, Queen's University,
                Kingston ON K7L 3N6 Canada}
\affiliation{Institute for Condensed Matter Physics, Lviv, Ukraine}
\author{R. J. Gooding}
\affiliation{Department of Physics, Queen's University,
                Kingston ON K7L 3N6 Canada}
%\thanks{This work was partially supported by NSERC of Canada and NATO.}


\date{\today}

\begin{abstract}
Motivated by recent experiments on undoped \LCO, which found pronounced
temperature-dependent anisotropies in the low-field magnetic susceptibility,
we have investigated a two-dimensional square lattice of $S=1/2$ spins
that interact via Heisenberg exchange plus the symmetric and anti-symmetric
Dzyaloshinskii-Moriya anisotropies. We describe the transition to
a state with long-ranged order, and find the spin-wave excitations,
with a mean-field theory, linear spin-wave analysis, and using Tyablikov's
RPA decoupling scheme. We find the different components of the susceptibility
within all of these approximations, both below and above the N\'eel temperature,
and obtain evidence of strong quantum fluctuations and spin-wave interactions
in a broad temperature region near the transition.

\end{abstract}

%\pacs{}
%\keywords{}

\maketitle

\section{Introduction}

Quantum magnetism of low-dimensional systems has attracted considerable attention
in recent years, in part due to the strong interest in the cuprate superconductors.
For example, it has been postulated that a strong antiferromagnetic (AF) exchange
interaction may be responsible for the high-temperature superconductivity
in these compounds.\cite{MMP}

The ubiquitous structural and electronic constituent of this latter class of materials
is the CuO$_2$ plane, and in this paper we consider the magnetic properties of
such planes in their undoped state. In particular, we consider the temperature
dependence of the static, uniform magnetic susceptibility for a single plane
in an undoped \LCO~crystal. This system is known to be an AF insulator
with a very simple structure, namely it can be approximately thought of as one
CuO$_2$ plane stacked between LaO planes, with this structural
unit repeated, in a body-centred tetragonal pattern, throughout all space.
However, a small orthorhombic distortion
introduces important spin-orbit couplings into the magnetic Hamiltonian, leading
to an AF state with a weak canted ferromagnetic moment. These spin-orbit
interactions are central to the results presented in this paper.

As was known from the start of research on the cuprate superconductors, a
complete knowledge of the properties of the spin-$\frac 12$ quantum Heisenberg AF
on a square lattice is an absolute necessity.\cite{ManousakisRMP} However, some experiments
have demonstrated that a complete description of the magnetic behaviour found in, {\emph{e.g.}}
\LCO, requires additional physics. Examples include
(i) weak ferromagnetism  in the low-temperature orthorhombic (LTO) phase;\cite{Thio,Kastner}
(ii) spin wave gaps with in- and out-of-plane modes;\cite{Keimer} and perhaps most importantly,
(iii) the unusual anisotropy of the magnetic susceptibility observed by Lavrov, Ando,
Komiya and Tsukada.\cite{Ando} It was this latter experiment that led us to complete
a sequence of theoretical investigations on a model that should describe such a three-dimensional
array of such CuO$_2$ planes modelling \LCO, a structure similar to those found in many
cuprate superconductors.
This manuscript summarizes the first of these studies, that concerned with a single
CuO$_2$ plane, with this plane described by a near-neighbour Heisenberg model plus
spin-orbit couplings as embodied by the antisymmetric and symmetric Dzyaloshinskii-Moriya
(DM) interactions.\cite{Dzyaloshinskii,Moriya}

An important point needs to be raised to clarify the applicability of this work to
a real physical system, such as \LCO. Firstly, note that according to the Mermin-Wagner theorem
a two-dimensional (2D) system with a continuous symmetry cannot undergo a continuous
phase transition, at any nonzero temperature, to a state with true long-ranged order. However,
when one includes both the antisymmetric and symmetric DM interactions this symmetry is lifted,
and thus the model that we study in this paper will have a true phase transition to
an ordered phase at some nonzero temperature, which we shall label by $T_N$, in
analogy to the N\'eel ordering temperature of a pure antiferromagnet. So, the ordered phase
for our model \textit{of a single plane} will include a weak ferromagnetic canted
moment, as well as long-ranged AF order. Note that current estimates\cite{Stain} of another interaction present in the
physical \LCO~system, that being a very weak AF interlayer coupling which is usually denoted by
$J_\perp$, is that this energy scale is close to that of the DM interactions, and thus it is likely
that both this exchange and the DM interactions are roughly equally responsible for the observed
transition. This serves to emphasize that our study of a single plane is not expected to accurately
explain {\emph{all}} of the observed magnetic properties of \LCO; in fact, this work
stands alone as a theoretical study of an isolated plane, but it is of considerable interest
to learn which experimental data can and which data can not be explained by such a single-plane model.

We focus on the role of the DM interaction between the neighbouring spins in a CuO$_2$ plane.
This interaction arises from the orthorhombic distortion in \LCO~(which is associated with
the small tilt of the CuO$_6$ octahedra) together with the spin-orbit interaction.
The DM interaction leads to a small canting of the Cu spins out of the plane, so that the weak
ferromagnetic order appears in each CuO$_2$ plane, and subsequently allows for the
formation of 3D AF order. This allows one to observe a pronounced peak in zero-field magnetic
susceptibility,\cite{notation} $\chi^c(T)$, and the earliest work on the importance of
this interaction focused on DM physics. To be specific, Thio {\emph{et al.}}\cite{Thio,Thio1}
analyzed their susceptibility data using a Landau theory expanded to sixth order, for a 2D Heisenberg
antiferromagnet with interlayer coupling and the DM-generated terms. They obtained reasonable
fits of their theory to the susceptibility and field dependent magnetization data, and deduced
 parameters which characterized magnetic properties of the \LCO~system. As we shall
explain below, we believe that the necessity of incorporating such higher-order terms into
their fits is suggestive of the important role played by spin-wave interactions, a conclusion
consistent with the results presented in this, and our future manuscripts on this problem.

Investigations of the magnetic ground state of \LCO~were performed by several groups
of authors, usually within the framework of the linear spin-wave (SW) theory.
The calculations were based on an effective model Hamiltonian derived by Moriya's
perturbation theory\cite{Moriya} applied to Hubbard type Hamiltonians by taking into account
the spin-orbit coupling.
In the most general form, the effective spin Hamiltonian, in addition to the isotropic exchange
interaction, includes the above-mentioned antisymmetric and symmetric DM interactions. The first
microscopic derivation of the spin Hamiltonian was performed by Coffey, Rice, and Zhang;\cite{Coffey2}
they estimated the antisymmetric DM coupling constants and showed that when the DM vectors
alternate a net ferromagnetic moment may be generated in the ground state.
Shekhtman, Entin-Wohlman, and Aharony\cite{Aharony1} subsequently showed that the symmetric anisotropies
contribute to the magnetic energy in the same order as the antisymmetric DM anisotropy, and
can never be neglected.
Several groups\cite{Koshibae,Aharony2,Stain} reexamined the Moriya' s theory and found
expressions for the effective spin Hamiltonian which includes both types of anisotropies.
The linear SW theory applied to such models at $T=0$ allows one to obtain previously reported
values of the spin-wave gaps at the centre of the 2D Brillouin zone, as well as to estimate
the magnitudes of the anisotropic-exchange interactions.
However, a detailed consideration of the model with the antisymmetric and symmetric DM anisotropies
at nonzero temperatures is up to now absent from the literature.

A very rough and simple approximation which can be used to study the effective magnetic model
at finite temperatures is the mean field approximation (MFA).
The MFA ignores effects of fluctuations and correlations between the spins, hence,
it fails for $T$ near $T_N$ and gives no short-range order above the transition temperature.
At very low $T$ the noninteracting SW theory is useful, and it gives a successful prediction
of the energy of low-lying excited states, and correctly reproduces the dominant term in the
low-$T$ magnetization. But, it fails near the phase transition point.
To analyze the high temperature behaviour the $1/T$ expansion method can be employed.
But, since the \LCO~crystal ordering temperature is much smaller than
the magnitude of the superexchange interaction ($T_{\rm N}<<J$), the high-temperature
expansion (to the first few orders in $J/T$) is not able to discuss the temperature region of
interest, that is $T$ near the transition temperature.

In the present paper time we consider the 2D spin-$\frac 12$ anisotropic quantum Heisenberg
antiferromagnet over the entire temperature range including both the symmetric and
anti-symmetric DM interactions. We employ the technique of double-time
temperature-dependent Green's functions within the framework of the random-phase approximation (RPA).
The first time such a scheme was used was by Tyablikov,\cite{Tyablikov} and he applied
this formalism to the Heisenberg ferromagnet (the RPA for magnetic models is often referred to
as Tyablikov's decoupling approximation). This work was generalized by Liu\cite{Liu} to obtain the longitudinal
correlation function, and this latter study is important in the development presented in our paper.
The important feature of this technique is that it deals with the entire temperature region and is
in a good agreement with the SW theory at low-$T$, as well as with $1/T$ expansions at high-$T$.
In this paper, within such a scheme, we find the transition temperature at which long-range order
would be established for an isolated plane.
We obtain the excitation spectrum, sublattice magnetization and susceptibility tensor as function of temperature
and coupling constants.
We also employ the MFA and SW theories to compare results of all of these approximation schemes, and
note the essential differences between them.

 Of course, many investigations of the 2D spin-$\frac{1}{2}$ have been completed previous
to this work. We have already mentioned the most popular and simple methods to study spin
models, that is phenomenological Landau theory, linear SW theory, the MFA, and high-temperature
expansions. They yield an analytical description of a wide range of physical properties and are very
useful for practical purposes.
 At the same time the great progress in the understanding of the ground state, thermodynamic
properties, and spin dynamics of the Heisenberg magnets was made with the use of the newer and
more complicated analytical schemes.
 Arovas and Auerbach\cite{Arovas} used a path-integral formulation of the MFA theory within
the Schwinger-boson representation.
 This method corresponds to the large-$N$ limit of the generalized SU$(N)$ model;
however, various difficulties with this method have been discussed in the literature.\cite{sarker89,desilva02}
 Takahashi\cite{Takahashi} has formulated and successfully applied the so-called
modified SW theory to the
Heisenberg model which reproduced the results of conventional SW theory and is closely related to
the Schwinger-boson theory.
 For the one dimensional chain, Takahashi's modified SW theory yields very good agreement with Bethe ansatz
results, as well as for the 2D classical ferromagnet at low-$T$ (in that it agrees with Monte Carlo results).
A self-consistent SW theory that is based on the boson-pseudofermion representation, was
developed to study thermodynamics of 2D systems, and was also applied to $S\geq 1$ systems
with an Ising-anisotropy 2D magnets.\cite{Irkhin}
An important feature of all these methods is that they can be used to describe both the
ordered and disordered (\textit{i.e.} the case of no long-range order) states.

Other related work includes: (i) The fermion representation to perform a $1/N$ expansion was used
by Affleck and Marston,\cite{Affleck}
large-$S$ 2D Heisenberg antiferromagnet in the long-wavelength limit; and
(ii) based on the diagrammatic method for the spin operators, the thermodynamics and the longitudinal
spin dynamics of Heisenberg magnets were studied.\cite{Sherman,Izyumov}
However, the most note-worthy success in the investigation of this system is the work i
of Chakravarty {\it et al.}\cite{Chakravarty}, who used a renormalization-group
approach to the quantum non-linear $\sigma$ model, the latter of
which describes the low-$T$ behaviour of the 2d Heisenberg AF in the long-wavelength
limit.

As will become apparent below, the formalism that we have chosen to implement is
more appropriate for this problem than any of those listed above, {{\emph{or}} the
theories listed above are too complicated to invoke when one goes beyond the
2D spin-$\frac{1}{2}$ Heisenberg AF and includes spin-orbit couplings.

The above few paragraphs summarize theoretical efforts that were directed towards the understanding of
the 2D S=1/2 square lattice AF. The application of these and related work to describe the magnetic
properties of so-called single-layer cuprate superconductors, such as \LCO, has attracted the
attention of many theorists, and fortunately an extensive review of this work, written
by Johnston, already exists.\cite{johnston}
In this review\cite{johnston} one can find the comparison of the temperature dependence of the
magnetic susceptibility for an AF Heisenberg square lattice calculated by different
analytical methods and quantum Monte Carlo calculations, and, apart from the (post-review) data
given by Lavrov \textit{et. al.}\cite{Ando}, the application of the analytical predictions
together with the numerical results show very good fitting to the experimental data for the
different single-layer cuprate compounds.

 Our paper is organized as follows.
 In \S\ref{sec:Model} we present the model Hamiltonian that we will study, introduce a convenient coordinate
transformation with which it is simple to complete analytical calculations,
and then derive the transformation that relates the static uniform susceptibility in both
coordinate systems.
 In \S\ref{sec:MFA} we derive and describe the MFA results, and then in \S\ref{sec:RPA}
we present our derivations from applying the Tyablikov/Liu approach to our model Hamiltonian.
 In \S\ref{sec:Results} we present a detailed examination of numerical results that follow from our work,
including a comparison of MFA, RPA and SW theories.
 Finally, in \S\ref{sec:conclusions} we summarize our paper including a brief discussion of the remainder
of the work that we have completed on the full three-dimensional problem.

\section{Model and Definitions:}
\label{sec:Model}

\subsection{Model Hamiltonian and the initial representation}
\label{subsec:Model_IR}

We consider a model for the Cu spins that are present in the \CuO~planes
of a \LCO~crystal in the low-temperature orthorhombic (LTO) phase and employ
a square lattice with nearest-neighbour interactions described by the
following effective magnetic Hamiltonian:\cite{Koshibae,Aharony1}
\be
\label{eq:H_DM}
    H=J\sum_{\langle i,j\rangle}{\bf S}_i\cdot{\bf S}_j +
    \sum_{\langle i,j\rangle}{\bf D}_{ij}\cdot({\bf S}_i\times{\bf S}_j)+
    \sum_{\langle i,j\rangle}{\bf S}_i\cdot
    \overleftrightarrow{\Gamma}_{ij}\cdot{\bf S}_j.
\ee
This Hamiltonian consists of the superexchange interaction
together with the antisymmetric Dzyaloshinskii-Moriya (DM)
interaction ($\bf D$ term) and the symmetric pseudodipolar
interaction ($\overleftrightarrow{\Gamma}$ term). As was discussed
in the introduction, the DM and pseudodipolar anisotropies arise
as a result of the mixture of Hubbard-type interaction energies
and spin-orbit coupling in the low symmetry crystal structure.

For the LTO phase, we use anisotropic interactions given by of the following form
\bea
\label{eq:DM}
    {\bf D}_{ab}=\frac d{\sqrt{2}}(-1,1,0),\;\;
    {\bf D}_{ac}=\frac d{\sqrt{2}}(-1,-1,0),
\eea
and
\bea
\label{eq:Gamma}
\overleftrightarrow{\Gamma}_{ab}{=}
  \left( \begin{array}{ccc}
     \Gamma_1 & \Gamma_2 & 0 \\
     \Gamma_2 & \Gamma_1 & 0 \\
     0 & 0 & \Gamma_3
         \end{array} \right)\!\!,\;
\overleftrightarrow{\Gamma}_{ac}{=}
  \left( \begin{array}{ccc}
     \Gamma_1 & -\Gamma_2 & 0 \\
     -\Gamma_2 & \Gamma_1 & 0 \\
     0 & 0 & \Gamma_3
         \end{array} \right)\!\!,
\eea
where the corresponding coordinates, in what we refer to as the ``initial representation"
in the LTO phase, are shown in Fig.~\ref{fig:vectors}(a). Note that the DM vector given
in Eq.~(\ref{eq:DM}) alternates in sign on successive bonds in the $a-b$
and in the $a-c$ direction of the lattice, as is represented schematically by the
double arrows in Fig.~\ref{fig:vectors}(b).
\begin{figure}[h]
\epsfxsize 0.45\textwidth\epsfbox{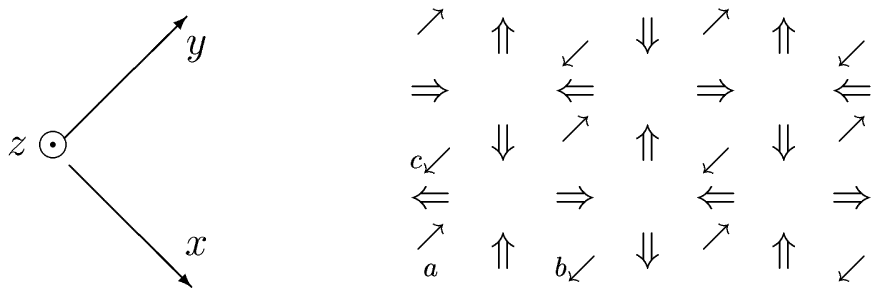}
%\newline
\hspace*{-.5cm}(a)\hspace*{4.8cm}(b)
\caption{\label{fig:vectors} (a) Coordinates in the initial representation.
                             (b) Thin arrows --- the Cu spins, open arrows --- the DM vectors.}
\end{figure}

We mention that the symmetric tensor $\overleftrightarrow{\Gamma}$ has been obtained
by several authors\cite{Aharony1,Bonesteel,Koshibae,Aharony2,Stain} in different forms.
We have chosen the general form of this tensor, from which other specialized
choices can be extracted. For instance, the form of the
symmetric tensor obtained by Koshibae, Ohta, and Maekawa\cite{Koshibae}
can be recovered from this definition if $\Gamma_3=\Gamma_2-\Gamma_1$.

In the LTO phase the classical ground state is determined uniquely,\cite{Koshibae,Aharony2}
and below the N\'eel temperature the Cu spin structure shows long-range antiferromagnetic order
with weak ferromagnetism (\textit{viz.} all spins cant out of the plane).
 To be concrete, in the classical ground state the spins are canted from in-plane antiferromagnetic
order by a small angle given by
\be
\theta = \frac 12\tan^{-1}\Big(\frac{d/\sqrt{2}}{J+\frac 12(\Gamma_1+\Gamma_3)}\Big),
\ee
and each plane has a net ferromagnetic moment in the $z$ direction
perpendicular to the CuO$_2$ planes (weak ferromagnetism).

In the simplified case of the zero pseudodipolar interaction
($\overleftrightarrow{\Gamma}=0$) it was found\cite{Coffey1,Coffey2}
that the ground-state spin configuration exhibits the rotational symmetry
about the DM vector which is the origin of the Goldstone mode in
the spin-wave spectrum.
Since in this simplified case there is a continuous symmetry in the
ground state, the thermal fluctuations destroy the long-range order
for any $T>0$, according to the Mermin and Wagner theorem.\cite{merminwagner}
In the general case of the model Hamiltonian of Eq.~(\ref{eq:H_DM}), the continuous
symmetry no longer exists and the spin-wave spectrum is gapped in
the long wavelength limit $\bq = 0$.
Consequently, \textit{the effect of fluctuations is reduced}.
That is, the DM (${\bf D}\neq 0$) together with pseudodipolar
($ \overleftrightarrow{\Gamma}\neq 0$) interactions can give rise
to long-range order for low (but nonzero) temperatures even for the
purely two-dimensional case (T$_{\rm N}>0$), and
the Mermin and Wagner theorem does not preclude the
possibility of a nonzero sublattice magnetization for nonzero temperatures
in this general case. (Note that his does not imply that the transition
to 3d long-ranged magnetic order is not influenced by the inter-planar exchange
coupling, but simply that this latter coupling is not, in general, necessary
to achieve such order.)

%As it was reported, for instance, by Koshibae, Ohta, and Maekawa
%\cite{Koshibae} the spectrum of the spin-wave excitations  of the
%model (\ref{eq:H_DM}) is always gapped.
%They showed the gap in the in-plane spin-wave mode is provided by
%the anisotropic-exchange interaction of superexchange mechanism
%and the out-of-plane spin-wave gap is provided by the
%anisotropic-exchange interaction of directexchange mechanism.

\subsection{Characteristic representation}
\label{subsec:Model_CR}

In solving this system, it is more convenient (theoretically)
to transform from the initial representation, given above, to the
characteristic representation (CR) in which the quantization axis
($z$) is in the direction of a classical moment characterizing
the ground state. In the present case there are two such classical
vectors in the direction of the canted moments (recall that we
are considering only a single \CuO~plane). Therefore, we introduce
two rotated coordinate systems, as shown in Fig.~\ref{fig:lattice}.
Spin degrees of freedom in the initial representation are denoted
by $\{S_i\}$, but in the characteristic representation we use
$\{\sigma_i\}$.
(We follow the notation that $i$-sites belong to sublattice 1,
whereas $j$-sites belong to sublattice 2.)
For the sites of sublattice 1 we apply a transformation of the form
\bw
\bea
\nonumber
\left (
 \begin{array}{c}
        \sigma_i^x\\
        \sigma_i^y\\
        \sigma_i^z\\
 \end{array}
\right )
&=&\frac 12\left (
 \begin{array}{ccc}
   \sin\theta{+}1 &\sin\theta{-}1 &{-}\sqrt{2}\cos\theta\\
   \sin\theta{-}1 &\sin\theta{+}1 &{-}\sqrt{2}\cos\theta\\
   \sqrt{2}\cos\theta &\sqrt{2}\cos\theta & 2\sin\theta\\
  \end{array}
 \right )
\left (
 \begin{array}{ccc}
   \frac 1{\sqrt{2}} & \frac {1}{\sqrt{2}} & 0 \\
   \frac {-1}{\sqrt{2}} &  \frac 1{\sqrt{2}} & 0 \\
   0&   0 & 1\\
  \end{array}
 \right )
\left (
 \begin{array}{c}
        S_i^x\\
        S_i^y\\
        S_i^z\\
 \end{array}
\right )\\
\label{eq:1_rot}
&=&\frac 1{\sqrt{2}}
\left (
 \begin{array}{ccc}
   1 &\sin\theta &-\cos\theta\\
  -1 &\sin\theta &-\cos\theta\\
  0&\sqrt{2}\cos\theta &\sqrt{2}\sin\theta\\
  \end{array}
 \right )
\left (
 \begin{array}{c}
        S_i^x\\
        S_i^y\\
        S_i^z\\
 \end{array}
\right )\!\!,
\end{eqnarray}
\ew
and for sublattice 2
\begin{eqnarray}
\label{eq:2_rot}
\left (
 \begin{array}{c}
        \sigma_j^x\\
        \sigma_j^y\\
        \sigma_j^z\\
 \end{array}
\right )=\frac 1{\sqrt{2}}
\left (
 \begin{array}{ccc}
   1 &\sin\theta &\cos\theta\\
  -1 &\sin\theta &\cos\theta\\
  0&-\sqrt{2}\cos\theta &\sqrt{2}\sin\theta\\
  \end{array}
 \right )\!\!\!
\left (
 \begin{array}{c}
        S_j^x\\
        S_j^y\\
        S_j^z\\
 \end{array}
\right )\!\!.
\eea
The quantization axes ($z$) of the new spin operators $\sigma_i$ and $\sigma_j$
coincide with the unit vectors in the direction of canted moments
moments Fig.~\ref{fig:lattice}.
\begin{figure}[h]
\epsfxsize 0.45\textwidth\epsfbox{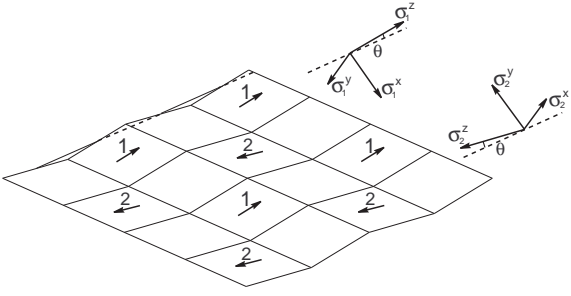}
\caption{\label{fig:lattice}Numbered arrows represent the Cu spin structure in a CuO$_2$ plane.
         Two sublattices 1 and 2 are introduced.
         For each sublattice the spin coordinate system
         within the characteristic representation ({\it i.e.}
         after the transformations given by (\ref{eq:1_rot}) and (\ref{eq:2_rot}))
         is shown.
         The thin net is shown only to simplify the visualization of the spin
         structure.}
\end{figure}

The model Hamiltonian of Eq.~(\ref{eq:H_DM}) in terms of the new operators $\sigma$ reads
\bea
\nonumber
H_{\rm CR}=\sum_{\langle i,j\rangle_{ab}}
    \left\{ A(\sigma^+_i\sigma^-_j+\sigma^-_i\sigma^+_j)\right.
    &&\!\!\!\!\!\!-B^*\sigma^+_i\sigma^+_j\\
\nonumber
    &&\!\!\!\!\!\!\left.-B\sigma^-_i\sigma^-_j
    -J_2\sigma^z_i\sigma^z_j
    \right\}\\
\nonumber
&&\hspace*{-4.6cm}+\sum_{\langle i,j\rangle_{ac}}
    \left\{ A(\sigma^+_i\sigma^-_j+\sigma^-_i\sigma^+_j)\right.
    +B\sigma^+_i\sigma^+_j\\
\nonumber
    &&\!\!\!\!\!\!\left.+B^*\sigma^-_i\sigma^-_j
    -J_2\sigma^z_i\sigma^z_j
    \right\},\\
\label{eq:H_CR}
\eea
where we introduced the following definitions
\bea
\nonumber
   && J_1=J+\Gamma_1,\\
\nonumber
   && J_2=\frac 12(\Gamma_1-\Gamma_3)+\sqrt{(d^2/2)+[J+\frac 12(\Gamma_1+\Gamma_3)]^2},\\
\nonumber
   && J_3=-\frac 12(\Gamma_1-\Gamma_3)+\sqrt{(d^2/2)+[J+\frac 12(\Gamma_1+\Gamma_3)]^2},\\
\nonumber
   && J_4=-\Gamma_2\sin\theta+\frac d{\sqrt{2}}\cos\theta,\\
\label{eq:JJJJ}\\
\label{eq:AB}
   && A = \frac{J_1-J_3}4,\qquad
   B = \frac{J_4}2+{\rm i} \frac{J_1+J_3}4.
\eea
The subscripts $\langle i,j\rangle_{ab}$ and $\langle i,j\rangle_{ac}$
in the summations of Eq.~(\ref{eq:H_CR}) imply the nearest neighbours in the
$ab$ and $ac$ directions, as shown in Fig.~\ref{fig:vectors}(b).

The form of the Hamiltonian in the characteristic representation is similar
to an XYZ model, but is clearly more complicated since terms of the
form $\sigma_i^-\sigma_j^--\sigma_i^+\sigma_j^+$ are present, which thus imply terms like
$S^x_iS^y_j$. Thus, we can extract from our results, in this representation,
the magnetic susceptibility of the XYZ calculated in both the mean-field
and random phase approximations, by setting the imaginary part $B=0$. We will consider numerical
results for this simpler model in a future publication.

\subsection{Magnetic susceptibility in the initial and characteristic representations}
\label{subsec:Mag_IRnCR}

We consider the response of the system, described by the Hamiltonian
$H$ in either the initial (Eq.~(\ref{eq:H_DM})) or
characteristic representation (Eq.~(\ref{eq:H_CR})),
to an externally applied constant magnetic field ${\bf h}$. It is convenient to
consider the application of this field in one direction only, which we take
to be the $\alpha$ direction \textit{of the initial representation},
\be
\label{eq:H-prime-IR}
H'=H-h^{\alpha}\sum^N_{l=1} S^{\alpha}_l,
\ee
where $\alpha = x$ or $y$ or $z$, it is to be noted that $\alpha$ is not summer over in
Eq.~(\ref{eq:H-prime-IR}), and $N$ is the number of the lattice sites.

The statistical operator of the system is required to evaluate ensemble averages
of relevant physical quantities, notably correlators and thermal Green's functions,
and can be written as
\be
\label{eq:stat_operator}
\rho = {\rm e}^{-\beta H'}={\rm e}^{-\beta H}
      T_{\tau}\exp\left\{h^{\alpha}\sum^N_{l=1}
      \int^{\beta}_0\!\! S^{\alpha}_l(\tau){\rm d}\tau\right\},
\ee
where $S_l(\tau)={\rm e}^{H\tau}S_l{\rm e}^{-H\tau}$ is the
operator in the Heisenberg representation for imaginary
time argument $\tau$, and $T_\tau$ is the time-ordering operator.
The zero-field susceptibility describes the response of the system to such a
field, and is defined to be
\be
\label{eq:chi_def1}
\chi^{\alpha}\equiv\frac{\partial\langle M^{\alpha}\rangle}{\partial h^{\alpha}}
          \Big|_{h^{\alpha}=0}=
          \frac 1N\sum^N_{l=1}\sum^N_{l'=1}
      \int^{\beta}_0\langle T_{\tau}S^{\alpha}_l(\tau)S^{\alpha}_{l'}(0)\rangle {\rm d}\tau,
\ee
where
\be
\label{eq:mag_moment}
\langle M^{\alpha}\rangle~=~ 1/N \sum^N_{l} \langle S^{\alpha}_l \rangle,
\ee
with correlators such as $\langle T_{\tau}S^{\alpha}_l(\tau)S^{\alpha}_{l'}(0)\rangle$
taken with respect to the zero field Hamiltonian $H$.

The square lattice is bipartite and can be divided into sublattices 1 and 2.
Then, by using the definitions
\bea
\label{eq:chi_def2}
\hspace*{-.7cm}
\chi^{\alpha}_{11}&=&\frac 2N\sum^{N/2}_{i=1}\sum^{N/2}_{i'=1}
          \int^{\beta}_0\!\!\langle T_{\tau}S^{\alpha}_i(\tau)S^{\alpha}_{i'}(0)\rangle {\rm d}\tau,\\
\label{eq:chi_def3}
\hspace*{-.7cm}
\chi^{\alpha}_{22}&=&\frac 2N\sum^{N/2}_{j=1}\sum^{N/2}_{j'=1}
          \int^{\beta}_0\!\!\langle T_{\tau}S^{\alpha}_j(\tau)S^{\alpha}_{j'}(0)\rangle {\rm d}\tau,\\
\label{eq:chi_def4}
\hspace*{-.7cm}
\chi^{\alpha}_{12}&=&\frac 2N\sum^{N/2}_{i=1}\sum^{N/2}_{j=1}
          \int^{\beta}_0\!\!\langle T_{\tau}S^{\alpha}_i(\tau)S^{\alpha}_{j}(0)\rangle {\rm d}\tau,\\
\hspace*{-.7cm}
\chi^{\alpha}_{21}&=&\frac 2N\sum^{N/2}_{j=1}\sum^{N/2}_{i=1}
          \int^{\beta}_0\!\!\langle T_{\tau}S^{\alpha}_j(\tau)S^{\alpha}_{i}(0)\rangle {\rm d}\tau,
\eea
where
\[
i,i'\in\mbox{ sublattice 1},\quad
j,j'\in\mbox{ sublattice 2}
\]
we can express the quantity of interest, $\chi^{\alpha}$, as
\be
\label{eq:chi-sum1}
\chi^{\alpha}=\frac 12\{ \chi^{\alpha}_{11}+\chi^{\alpha}_{22}
                            +\chi^{\alpha}_{12}+\chi^{\alpha}_{21} \}.
\ee
Then, using symmetry equivalent this simplifies (see below)
the calculation of the zero-field susceptibility in the initial representation to
\be
\label{eq:chi-sum2}
\chi^{\alpha}=\chi^{\alpha}_{11}+\chi^{\alpha}_{12}.
\ee

 The simpler form of Eq.~(\ref{eq:H_CR}) \textit{vs.} Eq.~(\ref{eq:H_DM}) makes clear
that it is desirable to perform calculations first using the characteristic representation,
and to then transform back into the initial representation.
 To this end we require the relevant form of the susceptibility tensor in the
characteristic representation.
 To begin, let us perform transformations ${\bf S}_1={\cal A}{\boldsymbol \sigma}_1$,
${\bf S}_2={\cal B}{\boldsymbol \sigma}_2$  (${\cal A}=[a_{\alpha\alpha'}]$,
${\cal B}=[b_{\alpha\alpha'}]$) to the characteristic representation, such that
the analogue of Eq.~(\ref{eq:H-prime-IR}) is
\begin{eqnarray}
\nonumber
 H'=H_{\rm CR}
 &-&\sum_{i=1}^{N/2} (a_{\alpha x}\sigma^x_i+a_{\alpha y}\sigma^y_i+a_{\alpha z}\sigma^z_i)h^{\alpha}_1\\
\label{eq:H-prime-CR}
 &-&\sum_{j=1}^{N/2} (b_{\alpha x}\sigma^x_j+b_{\alpha y}\sigma^y_j+b_{\alpha z}\sigma^z_j)h^{\alpha}_2.
\end{eqnarray}
Note that we have generalized the applied field to be ${\bf h_1}$ for sublattice 1,
and ${\bf h_2}$ for sublattice 2, and in general we will treat these as two
independent applied fields.
If we define the components of susceptibility in the characteristic representation as
\begin{eqnarray}
\label{eq:chi_def5}
\hspace*{-.3cm}
\chi^{\sigma^{\alpha}\sigma^{\alpha'}}_{11}&\!\!\!=\!\!\!&
    \frac 2N\sum^{N/2}_{i=1}\sum^{N/2}_{i'=1}\!
    \int^{\beta}_0\!\!\!\langle T_{\tau}\sigma_i^{\alpha}(\tau)
     \sigma_{i'}^{\alpha'}(0)\rangle{\rm d}\tau,\quad\\
\label{eq:chi_def6}
\hspace*{-.3cm}
\chi^{\sigma^{\alpha}\sigma^{\alpha'}}_{12}&\!\!\!=\!\!\!&
    \frac 2N\sum^{N/2}_{i=1}\sum^{N/2}_{j=1}\!
    \int^{\beta}_0\!\!\!\langle T_{\tau}\sigma_i^{\alpha}(\tau)
    \sigma_{j}^{\alpha'}(0)\rangle{\rm d}\tau,
\end{eqnarray}
then the susceptibility given in Eq.~(\ref{eq:chi_def2})
(\textit{N.B.} in the initial representation) can be written as
\bea
\nonumber
\hspace*{-0.5cm}
\chi^{\alpha}_{11}&=&\frac 2N\sum^{N/2}_{i=1}
                   \frac{\partial\langle S^{\alpha}_i\rangle}{\partial h^{\alpha}_1}\\
\nonumber
&=& \frac 2N\!\!\sum^{N/2}_{i=1}\!\!
 \left\{ a_{\alpha x}\frac{\partial\langle \sigma^x_i\rangle}{\partial h^{\alpha}_1}
        +a_{\alpha y}\frac{\partial\langle \sigma^y_i\rangle}{\partial h^{\alpha}_1}
 +a_{\alpha z}\frac{\partial\langle \sigma^z_i\rangle}{\partial h^{\alpha}_1} \right\}\\
\nonumber
&=& a_{\alpha x}^2\chi^{\sigma^x\sigma^x}_{11}
    +a_{\alpha x}a_{\alpha y}\chi^{\sigma^x\sigma^y}_{11}
    +a_{\alpha x}a_{\alpha z}\chi^{\sigma^x\sigma^z}_{11}\\
\nonumber
&+&a_{\alpha y}^2\chi^{\sigma^y\sigma^y}_{11}
    +a_{\alpha y}a_{\alpha x}\chi^{\sigma^y\sigma^x}_{11}
    +a_{\alpha y}a_{\alpha z}\chi^{\sigma^y\sigma^z}_{11}\\
\label{eq:chi_11}
&+&a_{\alpha z}^2\chi^{\sigma^z\sigma^z}_{11}
    +a_{\alpha z}a_{\alpha x}\chi^{\sigma^z\sigma^x}_{11}
    +a_{\alpha z}a_{\alpha y}\chi^{\sigma^z\sigma^y}_{11}\!\!,
\eea
and, in the same way (see Eq.~(\ref{eq:chi-sum2}))
\bea
\nonumber
\chi^{\alpha}_{12} &\!\!\!=\!\!\!& a_{\alpha x}b_{\alpha x}\chi^{\sigma^x\sigma^x}_{12}
    {+}a_{\alpha x}b_{\alpha y}\chi^{\sigma^x\sigma^y}_{12}
    {+}a_{\alpha x}b_{\alpha z}\chi^{\sigma^x\sigma^z}_{12}\\
\nonumber
 &\!\!\!+\!\!\!& a_{\alpha y}b_{\alpha y}\chi^{\sigma^y\sigma^y}_{12}
    {+}a_{\alpha y}b_{\alpha x}\chi^{\sigma^y\sigma^x}_{12}
    {+}a_{\alpha y}b_{\alpha z}\chi^{\sigma^y\sigma^z}_{12}\\
\label{eq:chi_12}
 &\!\!\!+\!\!\!& a_{\alpha z}b_{\alpha z}\chi^{\sigma^z\sigma^z}_{12}
    {+}a_{\alpha z}b_{\alpha x}\chi^{\sigma^z\sigma^x}_{12}
    {+}a_{\alpha z}b_{\alpha y}\chi^{\sigma^z\sigma^y}_{12}\!\!.
\eea

 The quantities $\chi^{\sigma^{\alpha}\sigma^{\alpha'}}$ that
are introduced above in Eqs.~(\ref{eq:chi_def5},\ref{eq:chi_def6})
have the following interpretation.
 For instance, the component $\chi^{\sigma^{\alpha}\sigma^{\alpha'}}_{12}$
determine the response of the expectation value
$2/N\sum^{N/2}_{i=1}\langle\sigma_i^{\alpha}\rangle$
of the spins of sublattice 1 to the magnetic field applied to the
spins sublattice 2 (no field applied to the spins of sublattice 1)
in the $\alpha'$ direction.  Indeed, the perturbation
$H'=H-h^{\alpha'}_2\sum^{N/2}_{j=1}\sigma^{\alpha'}_j$ formally
leads to the response
\bw
\be
\label{eq:iterpret1}
\frac 2N\frac {\partial}{\partial h_2^{\alpha'}}\sum^{N/2}_{i=1}\langle\sigma_i^{\alpha}\rangle=
\frac 2N\sum^{N/2}_{i=1}\sum^{N/2}_{j=1}
 \int^{\beta}_0\!\!\langle T_{\tau}\sigma_i^{\alpha}(\tau)
 \sigma_j^{\alpha'}(0)\rangle{\rm d}\tau
\equiv\chi^{\sigma^{\alpha}\sigma^{\alpha'}}_{12}.
\ee
Similarly, the response of the spins of sublattice 1 to the perturbation
$H'=H-h^{\alpha'}_1\sum^{N/2}_{i=1}\sigma^{\alpha'}_i$ is given by
\be
\label{eq:iterpret2}
 \frac 2N\frac {\partial}{\partial h_1^{\alpha'}}\sum^{N/2}_{i=1}\langle\sigma_i^{\alpha}\rangle=
 \frac 2N\sum^{N/2}_{i=1}\sum^{N/2}_{i'=1}
 \int^{\beta}_0\!\!\langle T_{\tau}\sigma_i^{\alpha}(\tau)
  \sigma_{i'}^{\alpha'}(0)\rangle{\rm d}\tau
  \equiv\chi^{\sigma^{\alpha}\sigma^{\alpha'}}_{11}.
\ee
\ew

So, by substituting the inverse to the CR transformation, given by
Eqs.~(\ref{eq:1_rot},\ref{eq:2_rot}), into Eqs.~(\ref{eq:chi_11},\ref{eq:chi_12}),
and taking into account that $\chi^{\sigma^x\sigma^z}{=}\chi^{\sigma^y\sigma^z}{=}
\chi^{\sigma^z\sigma^x}{=}\chi^{\sigma^z\sigma^y}{=}0$ in the characteristic
representation (which can be derived analytically), one obtains the desired transformation
between the two representations, namely
\bw
\bea
\label{eq:CRtoINx}
\chi^{x}= \chi^{x}_{11}+\chi^{x}_{12}&=&
          \frac 12(\chi^{\sigma^x\sigma^x}_{11}{+}\chi^{\sigma^x\sigma^x}_{12}
              {+}\chi^{\sigma^y\sigma^y}_{11}{+}\chi^{\sigma^y\sigma^y}_{12}
          {-}\chi^{\sigma^x\sigma^y}_{11}{-}\chi^{\sigma^x\sigma^y}_{12}
          {-}\chi^{\sigma^y\sigma^x}_{11}{-}\chi^{\sigma^y\sigma^x}_{12}
          ),\\
\chi^{y}= \chi^{y}_{11}+\chi^{y}_{12}&=&\frac{\sin^2(\theta)}{2}(
            \chi^{\sigma^x\sigma^x}_{11}{+}\chi^{\sigma^x\sigma^x}_{12}{+}
            \chi^{\sigma^y\sigma^y}_{11}{+}\chi^{\sigma^y\sigma^y}_{12}{+}
        \chi^{\sigma^x\sigma^y}_{11}{+}\chi^{\sigma^x\sigma^y}_{12}{+}
        \chi^{\sigma^y\sigma^x}_{11}{+}\chi^{\sigma^y\sigma^x}_{12})
\nonumber\\
\label{eq:CRtoINy}
       &&+\cos^2(\theta)(\chi^{\sigma^z\sigma^z}_{11}{-}\chi^{\sigma^z\sigma^z}_{12}),\\
\chi^{z}= \chi^{z}_{11}+\chi^{z}_{12}&=&\frac{\cos^2(\theta)}{2}(
            \chi^{\sigma^x\sigma^x}_{11}{-}\chi^{\sigma^x\sigma^x}_{12}{+}
            \chi^{\sigma^y\sigma^y}_{11}{-}\chi^{\sigma^y\sigma^y}_{12}{+}
        \chi^{\sigma^x\sigma^y}_{11}{-}\chi^{\sigma^x\sigma^y}_{12}{+}
        \chi^{\sigma^y\sigma^x}_{11}{-}\chi^{\sigma^y\sigma^x}_{12})
\nonumber\\
\label{eq:CRtoINz}
       &&+\sin^2(\theta)(\chi^{\sigma^z\sigma^z}_{11}{+}\chi^{\sigma^z\sigma^z}_{12}).
\eea
\ew

\section{Mean Field Analysis}
\label{sec:MFA}

In this section we develop the mean field approximation (MFA) for
the system defined by Eq.~(\ref{eq:H_DM}), and obtain the behaviour
of the magnetic susceptibility and a defining equation for the order
parameter as a function of temperature. In part
we include this derivation to make evident how the formalism
of \S~{\ref{subsec:Mag_IRnCR}} is applied to extract the zero-field
uniform magnetic susceptibility. However, and more importantly,
we will show that when the canting angle induced by the DM couplings
is small, there are significant deviations from the mean-field results,
{\emph{viz.}} quantum fluctuation effects are large. Thus, here
we establish the MFA susceptibility with which to make these comparisons.

Within the MFA we focus on one of the spins and replace its interaction
with other spins by an effective field. To this end the following
replacement is used:
\be
\label{eq:MFA_decoupling}
S_i^aS_j^b =
\langle S_i^a\rangle ~ S_j^b ~+~
S_i^a ~ \langle S_j^b\rangle ~-~
\langle S_i^a\rangle ~ \langle S_j^b\rangle,
\ee
where $a$ and $b$ can be equal to any of $x,y,z$.
It is to be noted that it is more convenient to perform the MFA calculations
starting from the model in the characteristic representation, and
thus we consider Eq.~(\ref{eq:H_CR}) and the analogue of the
above equation for the ${\bf \sigma}$ operators.

First, we find the equation for the order parameter.
The Hamiltonian Eq.~(\ref{eq:H_CR}) within the MFA reads as
\be
\label{eq:H_MFA}
H^{~MFA}_i = -{\cal Z}J_2\langle\sigma^z\rangle\sigma^z_i,
\ee
and we find that the order parameter, to be denoted by $\eta$, is found from
the solution of
\be
\label{eq:sigma_MFA}
\eta\equiv\langle\sigma^z\rangle = \frac 12 \tanh \left\{\frac{\beta}2
                              {\cal Z}J_2\langle\sigma^z\rangle\right\},
\ee
where $J_2$ is given by Eq.~(\ref{eq:JJJJ}), and ${\cal Z}$ is the coordination
number. From this equation it is immediately seen that within the MFA the N\'eel
temperature at which $\eta$ vanishes is
\be
\label{eq:T_N^MFA}
T^{MFA}_{N}=J_2=\frac 12(\Gamma_1{-}\Gamma_3)
+\sqrt{(d^2/2)+[J{+}\frac 12(\Gamma_1{+}\Gamma_3)]^2}
\ee

Now, we find the susceptibility of the system within the MFA below $T^{MFA}_N$.
First, we apply a magnetic field in the $z$ direction of the sublattice~1
\be
\label{eq:H-prime-MFA-withh1-compact}
H' = H - h^z_1\sum_i\sigma^z_i,
\quad i\mbox{-sites}\in 1\mbox{ sublattice}.
\ee
The Hamiltonian within the MFA can be written as
\be
\label{eq:H-prime-MFA-withh1}
H'^{~MFA}=-\sum_i\!\bigg(\!\sum_{\langle j\rangle_i}J_2\langle\sigma^z_j\rangle
             {+}h^z_1\bigg)\sigma^z_i
         -\sum_j\sum_{\langle i\rangle_j}
        J_2\langle\sigma^z_i\rangle\sigma^z_j,
\ee
where $\sum_{\langle i\rangle_j}$ means sum over all sites $i$ which are
nearest neighbours of site $j$. Then
\bea
\nonumber
\langle\sigma^z_i\rangle&=&\frac 12\tanh\left\{
         \frac{\beta}2\bigg(\sum_{\langle j\rangle_i} J_2\langle\sigma^z_j\rangle
          +h^z_1\bigg)\right\},\\
\nonumber
\langle\sigma^z_j\rangle&=&\frac 12\tanh\left\{
               \frac{\beta}2\sum_{\langle i\rangle_j}
           J_2\langle\sigma^z_i\rangle\right\}.\\
\label{eq:MFA_z1}
\eea
We write the mean value of $\sigma^z$ operators in the form
\be
\label{eq:define-delta-sigma}
\langle\sigma^z_i\rangle = \langle\sigma^z_1\rangle_0+\delta\sigma^z_1,\;
\langle\sigma^z_j\rangle = \langle\sigma^z_2\rangle_0+\delta\sigma^z_2,
\ee
where $\langle\sigma^z_1\rangle_0=\langle\sigma^z_2\rangle_0=\eta$,
is the expectation value of $\sigma^z$ operator in the absence of the field, and the
term $\delta\sigma^z$ is the part of $\langle\sigma^z\rangle$ induced
by the applied field. Since the applied field $h^z_1$ as well as the terms involving
$\delta\sigma^z$ are small, we may expand Eq.~(\ref{eq:MFA_z1}) in powers of
these terms. Then, we find
\bea
\nonumber
\chi_{11}^{\sigma^z\sigma^z}&\!\!\!=\!\!\!&\frac{\delta\sigma^z_1}{h^z_1}\Big|_{h^z_1=0}
     {=}   \frac{
           \frac{\beta}{4}{\rm sech}^2\left\{
           \frac{\beta}{2}{\cal Z}J_2\eta\right\}
     }
     {1{-}(\frac{\beta J_2{\cal Z}}{4})^2
         {\rm sech}^4\left\{\frac{\beta}{2}{\cal Z}J_2\eta\right\}
     },\\
\nonumber
\chi_{21}^{\sigma^z\sigma^z}&\!\!\!=\!\!\!&\frac{\delta\sigma^z_2}{h^z_1}\Big|_{h^z_1=0}
     {=}   \frac{
          {\cal Z}J_2(\frac{\beta}{4})^2
          {\rm sech}^4\left\{\frac{\beta}2 {\cal Z}J_2\eta\right\}
     }
     {1{-}(\frac{\beta J_2{\cal Z}}{4})^2
         {\rm sech}^4\left\{\frac{\beta}{2}{\cal Z}J_2\eta\right\}
     }.\\
\label{eq:MFA_z2}
\eea

Due to the complicated couplings found in Eq.~(\ref{eq:H_CR}), the transverse
components are much more involved to calculate.
Applying a field in the $x$ direction to the spins of sublattice 1 we consider
\be
H' = H - h^x_1\sum_i\sigma^x_i,
\quad i\mbox{-sites}\in 1\mbox{-sublattice},
\ee
and within the MFA we thus examine
\bea
\nonumber
 H'^{~MFA} =&\!\!-\!\!&\sum_i([{\mathfrak h}^x_1+h^x_1]\sigma^x_i+{\mathfrak h}^y_1\sigma^y_i
+{\mathfrak h}^z_1\sigma^z_i)\\
\label{eq:MFA_x1}
&\!\!-\!\!&\sum_j({\mathfrak h}^x_2\sigma^x_j+{\mathfrak h}^y_2\sigma^y_j
+{\mathfrak h}^z_2\sigma^z_j).
\eea
Similarly, by applying a field in the $y$ direction to the spins
of sublattice 1 we consider
\bea
\nonumber
 H'^{~MFA} =&\!\!-\!\!&\sum_i({\mathfrak h}^x_1\sigma^x_i+[{\mathfrak h}^y_1+h^y_1]\sigma^y_i
+{\mathfrak h}^z_1\sigma^z_i)\\
\label{eq:MFA_y1}
&\!\!-\!\!&\sum_j({\mathfrak h}^x_2\sigma^x_j+h^y_2\sigma^y_j
+{\mathfrak h}^z_2\sigma^z_j),
\eea
where
\bea
 \nonumber
 &&{\mathfrak h}^x_1 = \sum_{\langle j\rangle_i}\{-2A\langle\sigma^x_j\rangle
                                             +2\Im B\langle\sigma^y_j\rangle\},\\
\nonumber
 &&{\mathfrak h}^x_2 = \sum_{\langle i\rangle_j}\{-2A\langle\sigma^x_i\rangle
                                             +2\Im B\langle\sigma^y_i\rangle\},\\
 \nonumber
 &&{\mathfrak h}^y_1 = \sum_{\langle j\rangle_i}\{-2A\langle\sigma^y_j\rangle
                                             +2\Im B\langle\sigma^x_j\rangle\},\\
  \nonumber
  &&{\mathfrak h}^y_2 = \sum_{\langle i\rangle_j}\{-2A\langle\sigma^y_i\rangle
                                             +2\Im B\langle\sigma^x_i\rangle\},\\
 \nonumber
 &&{\mathfrak h}^z_1 = \sum_{\langle j\rangle_i}J_2\langle\sigma^z_j\rangle,\;\;
  {\mathfrak h}^z_2 = \sum_{\langle i\rangle_j}J_2\langle\sigma^z_i\rangle,
\eea
where $\Im B$ denotes the imaginary part of $B$.
Then, the system of equations determining the transverse components of susceptibility
Eq.~(\ref{eq:iterpret1}) and Eq.~(\ref{eq:iterpret2}) within the MFA scheme is found
to be
\bea
\nonumber
 -\frac {J_2}{2}\chi^{\sigma^x\sigma^x}_{11} &=& A\chi^{\sigma^x\sigma^x}_{21}
  -\Im B\chi^{\sigma^y\sigma^x}_{21}-\frac 1{2{\cal Z}},\\
\nonumber
 -\frac {J_2}{2}\chi^{\sigma^x\sigma^x}_{21} &=& A\chi^{\sigma^x\sigma^x}_{11}
  -\Im B\chi^{\sigma^y\sigma^x}_{11},\\
\nonumber
 -\frac {J_2}{2}\chi^{\sigma^y\sigma^x}_{11} &=& A\chi^{\sigma^y\sigma^x}_{21}
  -\Im B\chi^{\sigma^x\sigma^x}_{21},\\
\nonumber
 -\frac {J_2}{2}\chi^{\sigma^y\sigma^x}_{21} &=& A\chi^{\sigma^y\sigma^x}_{11}
  -\Im B\chi^{\sigma^x\sigma^x}_{11},\\
\nonumber
  -\frac {J_2}{2}\chi^{\sigma^x\sigma^y}_{11} &=& A\chi^{\sigma^x\sigma^y}_{21}
  -\Im B\chi^{\sigma^y\sigma^y}_{21},\\
\nonumber
  -\frac {J_2}{2}\chi^{\sigma^x\sigma^y}_{21} &=& A\chi^{\sigma^x\sigma^y}_{11}
  -\frac{\cal Z}{4}\Im B\chi^{\sigma^y\sigma^y}_{11},\\
\nonumber
  -\frac {J_2}{2}\chi^{\sigma^y\sigma^y}_{11} &=& A\chi^{\sigma^y\sigma^y}_{21}
  -\Im B\chi^{\sigma^x\sigma^y}_{21}-\frac 1{2{\cal Z}},\\
\nonumber
  -\frac {J_2}{2}\chi^{\sigma^y\sigma^y}_{21} &=& A\chi^{\sigma^y\sigma^y}_{11}
  -\Im B\chi^{\sigma^x\sigma^y}_{11}.\\
\label{eq:MFA_x3}
\eea

The solution of this systems, Eq.~(\ref{eq:MFA_x3}) turns out to be
\bea
\nonumber
 \chi^{\sigma^x\sigma^x}_{11}&\!\!=\!\!&\chi^{\sigma^x\sigma^x}_{22}=
 \chi^{\sigma^y\sigma^y}_{11}=\chi^{\sigma^y\sigma^y}_{22}=
                  \frac{J_2/2{+}A}{4{\cal Z}\omega^2_1}+
                  \frac{J_2/2{-}A}{4{\cal Z}\omega^2_2},\\
\nonumber
 \chi^{\sigma^x\sigma^x}_{12}&\!\!=\!\!&\chi^{\sigma^x\sigma^x}_{21}=
 \chi^{\sigma^y\sigma^y}_{12}=\chi^{\sigma^y\sigma^y}_{21}=
                  \frac{J_2/2{+}A}{\omega^2_1}-
                  \frac{J_2/2{-}A}{4{\cal Z}\omega^2_2},\\
\nonumber
 \chi^{\sigma^x\sigma^y}_{11}&\!\!=\!\!&\chi^{\sigma^x\sigma^y}_{22}=
 \chi^{\sigma^y\sigma^x}_{11}=\chi^{\sigma^y\sigma^x}_{22}=
 \frac{\Im B}{4{\cal Z}}\left(\frac 1{\omega^2_1}-\frac 1{\omega^2_2} \right),\\
\nonumber
 \chi^{\sigma^x\sigma^y}_{12}&\!\!=\!\!&\chi^{\sigma^x\sigma^y}_{21}=
 \chi^{\sigma^y\sigma^x}_{12}=\chi^{\sigma^y\sigma^x}_{21}=
 \frac{\Im B}{4{\cal Z}}\left(\frac 1{\omega^2_1}+\frac 1{\omega^2_2} \right),\\
\label{eq:all_chi_sigma}
\eea
where
\bea
\nonumber
\omega_1&=&\sqrt{(J_2/2+A)^2-{\Im B}^2},\\
\nonumber
\omega_2&=&\sqrt{(J_2/2-A)^2-{\Im B}^2}.\\
\label{eq:omega}
\eea
Using the relation between the components of susceptibility in the initial
and characteristic representations given in Eqs.~(\ref{eq:CRtoINx})-(\ref{eq:CRtoINz}),
we obtain the final result for zero-field
uniform susceptibility within the MFA below the MFA ordering temperature,
$T^{MFA}_N$, {\emph{viz.}}
\bea
\label{eq:MFA_SxSx}
\chi^{x~MFA}=&&\!\!\!\!\frac 14 \frac 1{J_1+J_2},\\
\nonumber
\chi^{y~MFA}=&&\!\!\!\!\frac 14 \frac {\sin^2(\theta)}{J_2-J_3}
           +\frac{\cos^2(\theta)}4
       \frac{{\rm sech}^2\left\{\frac{\beta}2 zJ_2\eta\right\}}
       {T+J_2~{\rm sech}^2\left\{\frac{\beta}2 zJ_2\eta\right\}},\\
\label{eq:MFA_SySy}\\
\nonumber
\chi^{z~MFA}=&&\frac 14 \frac {\cos^2(\theta)}{J_2+J_3}
           +\frac{\sin^2(\theta)}4
       \frac{{\rm sech}^2\left\{\frac{\beta}2 zJ_2\eta\right\}}
       {T-J_2~{\rm sech}^2\left\{\frac{\beta}2 zJ_2\eta\right\}},\\
\label{eq:MFA_SzSz}
\eea
with the equation for the order parameter $\eta$ given by Eq.~(\ref{eq:sigma_MFA}).
(For $d=\Gamma_i=0$, implying that $\theta=0$ and $J_2=J$,
the above seemingly complicated results indeed reduce to the
correct MFA expression for the susceptibility.)

The following comments on the MFA result are in order. First, note that
for physical values of $d$ and $\Gamma_i$ ($d,~\Gamma_i~\ll~J$) the canting
angle out of the $xy$ plane is very small; thus, since the AF moment
is in the $yz$ plane and nearly aligned along the $\pm y$ axes, $\chi^z$
diverges at $T^{MFA}_N$, but the other two components remain finite at
the transition. However, while the $x$ component of the susceptibility
remains independent of temperature, since the canting produces a net FM
moment in the $z$ direction that is coupled to the $y$ component of
the local moment, there is an additional increase of $\chi^y$ as the transition
is approached from below.

Now consider the paramagnetic temperature region ($T>T_{\rm N}$), for which the
only components with nonzero spin expectation values are those driven by the applied field.
Following similar considerations to above, the final results for the components of
susceptibility in the initial representation for high temperatures ($T>T_{\rm N}$) reads
\begin{eqnarray}
\label{eq:MFA_SxSx_para}
\chi^{x~ MFA}&=&\frac 14 \frac 1{J_1+T},\\
\label{eq:MFA_SySy_para}
\chi^{y~MFA}&=&\frac 14 \frac {\sin^2(\theta)}{T-J_3}
           +\frac 14 \frac{\cos^2(\theta)}{T+J_2},\\
\label{eq:MFA_SzSz_para}
\chi^{z~MFA}&=&\frac 14 \frac {\cos^2(\theta)}{T+J_3}
           +\frac 14 \frac{\sin^2(\theta)}{T-J_2}.
\end{eqnarray}
Note that in the limit $T\to T^{MFA}_{\rm N}=J_2$ we obtain that the $x,y$ components of
the susceptibility are continuous at the transition, whereas the $z$ component of
the susceptibility diverges at the N\'eel point, from above or below, owing to the
presence of the weak ferromagnetic moment that first develops at the transition.

\section{Linear response theory within the RPA}
\label{sec:RPA}

\subsection{Susceptibility below $T_N$}

In this section we derive expressions for the static, uniform susceptibility
within the RPA below the ordering temperature, $T_N$. Note that this
temperature is determined with the RPA, and is not equivalent to that
found in the previous section.

We employ thermal Green's functions in the analysis of
the spin Hamiltonian given in Eq.~(\ref{eq:H_DM}) with spin $\frac 12$.
The definition of such Green's functions for two Bose operators
$A$, $B$ and the corresponding equation of motion, are given by
\bea
\label{eq:defG}
\hspace*{-.7cm}
G_{AB}(\tau) &\!\!=\!\!& \langle T_\tau A(\tau)B(0)\rangle,\\
\hspace*{-.7cm}
\frac{{\rm d}G_{AB}(\tau)}{{\rm d}\tau}&\!\!=\!\!&\delta(\tau)\langle[A,B]\rangle
 + \langle T_\tau[H(\tau),A(\tau)]B(0)\rangle.
\eea

As discussed in the introduction, we adopt a procedure that was introduced
by Liu,\cite{Liu} as this technique allows for us to find longitudinal component
of the susceptibility. To this end, we introduce the perturbed Hamiltonian
(in the characteristic representation)
\be
\label{eq:H_pert}
H^f_1 = H_{\rm CR} - f \sum_i\sigma^z_i,
\ee
where $f$ is a small fictitious field; note that the field is applied to the
spins of {\emph {sublattice 1 only}}, and within the present paper we restrict $f$
to be constant and static.

In the imaginary-time formalism, the Green's functions to be used are
\begin{eqnarray}
\nonumber
 G^f_{ln}(\tau)&\!\!=\!\!&\langle T_{\tau}\sigma^+_l(\tau)\sigma^-_n(0)\rangle^f,\\
\nonumber
 G^{f-}_{ln}(\tau)&\!\!=\!\!&\langle T_{\tau}\sigma^-_l(\tau)\sigma^-_n(0)\rangle^f,\;
 l\! \in\!\! \mbox{ sublattice 1},\\
 \nonumber
 G^f_{n'n}(\tau)&\!\!=\!\!&\langle T_{\tau}\sigma^+_{n'}(\tau)\sigma^-_n(0)\rangle^f,\\
\nonumber
 G^{f-}_{n'n}(\tau)&\!\!=\!\!&\langle T_{\tau}\sigma^-_{n'}(\tau)\sigma^-_n(0)\rangle^f,\;
 n\! \in\!\! \mbox{ sublattice 2},\\
\label{eq:GGG}
\end{eqnarray}
where the expectation values are taken with  respect to the perturbed
Hamiltonian in Eq.~(\ref{eq:H_pert}). After an expansion in a power series of $f$ we
can write
\be
\label{eq:def_Gf}
 G^f_{ln}(\tau) = G^{(0)}_{ln}(\tau) + f G^{(1)}_{ln}(\tau) + O(f^2).
\ee
Since $G^{(0)}_{ln}(\tau)=G_{ln}(\tau)$, from now drop the superscript and use
\be
\label{eq:explanation1}
 G^f_{ln}(\tau) = G_{ln}(\tau) + f G^{(1)}_{ln}(\tau) + O(f^2).
\ee
Also, we introduce
\be
\label{eq:explanation2}
 \langle \sigma^z_i(\tau)\rangle^f =
   \langle \sigma^z_i\rangle + f{\rm v}_i+O(f^2),
\ee
where, due to the translation periodicity
$\langle \sigma^z_i\rangle=\eta$, the order parameter at $f=0$.

The equation of motion for the Green's function $G^f_{ln}(\tau)$ is given by
\begin{eqnarray}
\nonumber
\hspace*{-1.cm}\frac{{\rm d}G^f_{ln}(\tau)}{{\rm d}\tau}&\!\!=\!\!&2\delta(\tau)\delta_{ln}
  \langle \sigma^z_l\rangle^f\\
\label{eq:eqom_Gf}
\hspace*{-1.cm}&\!\!+\!\!&\langle T_\tau[H_{\rm CR}(\tau),\sigma^+_l(\tau)]\sigma^-_n(0)\rangle^f
  -fG^f_{ln}.
\end{eqnarray}
In order to solve this equation for the Green's function it must be linearized.
We will use the random phase approximation (RPA), in which the
fluctuations of $\sigma^z$ are ignored and the operator $\sigma^z$ is replaced
by its mean value $\langle\sigma^z\rangle^f$ --- this is the so-called Tyablikov's
decoupling.\cite{Tyablikov}
For example
\bea
 \nonumber
\hspace*{-1.4cm}
 &&\langle T_{\tau} \sigma^z_l(\tau)\sigma^+_i(\tau)\sigma^-_j(0) \rangle^f
 \to\\
\label{eq:Tyablikov_decoupling}
\hspace*{-1.4cm}
&&\quad\to\langle \sigma^z_l(\tau)\rangle^f
 \langle T_{\tau} \sigma^+_i(\tau)\sigma^-_j(0) \rangle^f
{=}\langle \sigma^z_l(\tau)\rangle^f G^f_{ij}(\tau).
\eea
After this decoupling is introduced, Eq.~(\ref{eq:eqom_Gf}) is found to be
\bw
\begin{eqnarray}
\nonumber
 \fl\fl
   \frac{{\rm d}G^f_{ln}(\tau)}{{\rm d}\tau}&\!\!=\!\!&
   2\delta(\tau)\delta_{ln}\langle \sigma^z_l\rangle^f
   -\sum_{\delta_{ab}}\left\{\!2\langle \sigma^z_l(\tau)\rangle^f
    [AG^f_{(l{+}\delta)n}(\tau){-}BG^{f-}_{(l{+}\delta)n}(\tau)]
    +J_2\langle \sigma^z_{l{+}\delta}(\tau)\rangle^fG^f_{ln}(\tau)\!\right\}\\
\label{eq:eqom_GF_2}
  &\!\!-\!\!\!&\sum_{\delta_{ac}}\left\{2\langle \sigma^z_l(\tau)\rangle^f
   [AG^f_{(l{+}\delta)n}(\tau){+}B^*G^{f-}_{(l{+}\delta)n}(\tau)]
   +J_2\langle \sigma^z_{l{+}\delta}(\tau)\rangle^fG^f_{ln}(\tau)\right\}
   -fG^f_{ln}(\tau),
\end{eqnarray}
\ew
where $\sum_{\delta_{ab}}$ refers to a summation over the nearest neighbours
of the site $l$ in the $ab$ direction, and similarly for  $\sum_{\delta_{ac}}$
 --- see Fig.~\ref{fig:vectors}(b). Here, all sites $l+\delta$ belong to the sublattice~2.

We introduce the Fourier transformation in the momentum-frequency representation
for the Green's function and the spin operator
\be
\label{eq:Fourier1}
 G^f_{ln}(\tau)=\frac 2{N\beta}\sum_{\bk, m}G^f_{12}(\bk,\omega_m)
  {\rm e}^{{\rm i}\bk\cdot({\bf R}_l-{\bf R}_n)}{\rm e}^{-{\rm i}\omega_m\tau}\!\!,
\ee
\bea
\nonumber
 \langle\sigma^z_l(\tau)\rangle^f&=&
   \frac 1{\beta}\sum_{\bk, m}\langle \sigma^z_1(\bk,\omega_m)\rangle^f
   {\rm e}^{-{\rm i}\bk\cdot{\bf R}_l}{\rm e}^{-{\rm i}\omega_m\tau}\\
\label{eq:Fourier2}
   &=& \sum_{\bk}\delta(\bk)[\eta + f{\rm v}_1]{\rm e}^{-{\rm i}\bk\cdot{\bf R}_l},
\eea
where the expansion in Eq.~(\ref{eq:explanation2}) and the linear response to the uniform
perturbation expressed by ${\rm v}_1(\bk)=\delta(\bk){\rm v}_1$ were taken into account.
In the transformation given by Eqs.~(\ref{eq:Fourier1},\ref{eq:Fourier2}), the sum over
$\bk$ runs over $\frac 12 N$ points of the first zone in the momentum space, and
$\omega_n = 2\pi n/\beta$ for $n\in\mathbb{Z}$ are the Bose Matsubara frequencies.
Then, we can write down the equation for the Green's function $G^f_{ln}(\tau)$
in the form
\bea
\nonumber
\hspace*{-0.5cm}
 -{\rm i}\omega_m G^f_{12}(\bk,\omega_m)=
   &\!\!-\!\!&fG^f_{12}(\bk,\omega_m)\\
\nonumber
\hspace*{-0.5cm}
   &\!\!-\!\!&{\cal Z}J_2[\eta + f{\rm v}_2]G^f_{12}(\bk,\omega_m)\\
\nonumber
\hspace*{-0.5cm}
   &\!\!-\!\!&2{\cal Z}A_{\bk}[\eta + f{\rm v}_1]G^f_{22}(\bk,\omega_m)\\
\label{eq:eq3}
\hspace*{-0.5cm}
   &\!\!+\!\!&2{\cal Z}B_{\bk}[\eta + f{\rm v}_1]G^{f-}_{22}(\bk,\omega_m),
\eea
where, as before, ${\cal Z}$ is the coordination number, and we introduce
\bea
 && A_{\bk} = A\gamma_{\bk},\quad
   B_{\bk} = (\Re B) \gamma_{\bk}'+(\Im B) \gamma_{\bk}, \\
\nonumber
 && \gamma_{\bk}=\frac 12(\cos k_x+\cos k_y),\quad
    \gamma'_{\bk}=\frac 12(\cos k_x-\cos k_y).
\eea
From these we can write down the following two equations:
\bea
  \frac {{\rm i}\omega_m}{2{\cal Z}\eta}G_{12}&=& \frac{J_2}2 G_{12}
  +A_{\bk}G_{22}-B_{\bk}G^-_{22},\\
\nonumber
   \frac {{\rm i}\omega_m}{2{\cal Z}\eta}G^{(1)}_{12}&=&\frac 1{2{\cal Z}\eta}G_{12}\\
\nonumber
   &&+\frac{{\rm v}_2}{\eta}\frac {J_2}{2}G_{12}+\frac {J_2}2G^{(1)}_{12}
   +\frac{{\rm v}_1}{\eta}A_{\bk}G_{22}+A_{\bk}G^{(1)}_{22}\\
 &&-\frac{{\rm v}_1}{\eta}B_{\bk}G^-_{22}-B_{\bk}G^{(1)-}_{22},
\eea
where in all equations we drop the wave vector and frequency dependencies
for the Green's functions, that is $G=G(\bk,\omega_m)$
and $G^{(1)}=G^{(1)}(\bk,\omega_m)$.

In the same way we obtain the equations of motion for the other Green's
functions (see Eq.~(\ref{eq:GGG})) within the RPA scheme.
The final systems of equations for zeroth- and first-order quantities can
be written as
\bw
\bea
\nonumber
   \left(\frac {{\rm i}\omega_m}{2{\cal Z}\eta}-\frac {J_2}2\right)
   G_{12}&=&A_{\bk}G_{22}
  -B_{\bk}G^-_{22},
\quad\qquad
   \left(\frac {{\rm i}\omega_m}{2{\cal Z}\eta}+\frac {J_2}2\right)
   G^-_{12}=-A_{\bk}G^-_{22}
  +B^*_{\bk}G_{22},\\
\nonumber
 \left(\frac {{\rm i}\omega_m}{2{\cal Z}\eta}-\frac {J_2}2\right)
  G_{22}&=&A_{\bk}G_{12}
  -B_{\bk}G^-_{12}-\frac 1{\cal Z},
  \quad
  \left(\frac {{\rm i}\omega_m}{2{\cal Z}\eta}+\frac {J_2}2\right)
   G^-_{22}=-A_{\bk}G^-_{12}
  +B^*_{\bk}G_{12};\\
\label{eq:Z_eq}
\eea
\bea
\nonumber
\hspace*{-1cm}
   \left(\frac {{\rm i}\omega_m}{2{\cal Z}\eta}-\frac {J_2}2\right)
   G^{(1)}_{12} &=&
   A_{\bk}G^{(1)}_{22} - B_{\bk}G^{(1)-}_{22}
   + \bigg\{ \frac{{\rm v}_2}{\eta}\frac {J_2}2
            +\frac{{\rm v}_1}{\eta} \bigg(\frac {{\rm i}\omega_m}{2{\cal Z}\eta}-\frac {J_2}2\bigg)
        +\frac 1{2{\cal Z}\eta}
        \bigg\}G_{12},\\
\hspace*{-1cm}\nonumber
  \left(\frac {{\rm i}\omega_m}{2{\cal Z} \eta}-\frac {J_2}2\right)
   G^{(1)}_{22}&=&
   A_{\bk}G^{(1)}_{12} - B_{\bk}G^{(1)-}_{12}
   + \bigg\{ \frac{{\rm v}_1}{\eta}\frac {J_2}2
            +\frac{{\rm v}_2}{\eta} \bigg(\frac {{\rm i}\omega_m}{2{\cal Z}\eta}-\frac {J_2}2\bigg)
            \bigg\}G_{22},\\
\hspace*{-1cm}\nonumber
   \left(\frac {{\rm i}\omega_m}{2{\cal Z}\eta}+\frac {J_2}2\right)
   G^{(1)-}_{12} &=&
   - A_{\bk}G^{(1)-}_{22} + B^*_{\bk}G^{(1)}_{22}
   - \bigg\{  \frac{{\rm v}_2}{\eta}\frac {J_2}2
            - \frac{{\rm v}_1}{\eta} \bigg(\frac {{\rm i}\omega_m}{2{\cal Z}\eta}+\frac {J_2}2\bigg)
        + \frac 1{2{\cal Z}\eta}
            \bigg\}G^-_{12},\\
\hspace*{-1cm}\nonumber
   \left(\frac {{\rm i}\omega_m}{2{\cal Z}\eta}+\frac {J_2}2\right)
   G^{(1)-}_{22} &=&
   - A_{\bk}G^{(1)-}_{12} + B^*_{\bk}G^{(1)}_{12}
   - \bigg\{ \frac{{\rm v}_1}{\eta}\frac {J_2}2
            -\frac{{\rm v}_2}{\eta} \bigg(\frac {{\rm i}\omega_m}{2{\cal Z}\eta}+\frac {J_2}2\bigg)
            \bigg\}G^-_{22};\\
\label{eq:F_eq}
\eea
\bea
\nonumber
\hspace*{-.7cm}
   \left(\frac {{\rm i}\omega_m}{2{\cal Z}\eta}-\frac {J_2}2\right)
   G^{(1)}_{21} &=&
   A_{\bk}G^{(1)}_{11} - B_{\bk}G^{(1)-}_{11}
   + \bigg\{ \frac{{\rm v}_1}{\eta}\frac {J_2}2
            +\frac{{\rm v}_2}{\eta} \bigg(\frac {{\rm i}\omega_m}{2{\cal Z}\eta}-\frac {J_2}2\bigg)
            \bigg\}G_{12},\\
\hspace*{-.7cm}\nonumber
  \left(\frac {{\rm i}\omega_m}{2{\cal Z}\eta}-\frac {J_2}2\right)
   G^{(1)}_{11}&=&
   A_{\bk}G^{(1)}_{21} - B_{\bk}G^{(1)-}_{21}
   + \bigg\{ \frac{{\rm v}_2}{\eta}\frac {J_2}2
            +\frac{{\rm v}_1}{\eta} \bigg(\frac {{\rm i}\omega_m}{2{\cal Z}\eta}-\frac {J_2}2\bigg)
        +\frac 1{2{\cal Z}\eta}
            \bigg\}G_{22},\\
\hspace*{-.7cm}\nonumber
   \left(\frac {{\rm i}\omega_m}{2{\cal Z}\eta}+\frac {J_2}2\right)
   G^{(1)-}_{21} &=&
   - A_{\bk}G^{(1)-}_{11} + B^*_{\bk}G^{(1)}_{11}
   - \bigg\{ \frac{{\rm v}_1}{\eta}\frac {J_2}2
            -\frac{{\rm v}_2}{\eta} \bigg(\frac {{\rm i}\omega_m}{2{\cal Z}\eta}+\frac {J_2}2\bigg)
            \bigg\}G^-_{12},\\
\hspace*{-.7cm}\nonumber
   \left(\frac {{\rm i}\omega_m}{2{\cal Z}\eta}+\frac {J_2}2\right)
   G^{(1)-}_{11} &=&
   - A_{\bk}G^{(1)-}_{21} + B^*_{\bk}G^{(1)}_{21}
   - \bigg\{ \frac{{\rm v}_2}{\eta}\frac {J_2}2
            -\frac{{\rm v}_1}{\eta} \bigg(\frac {{\rm i}\omega_m}{2{\cal Z}\eta}+\frac {J_2}2\bigg)
        +\frac 1{2{\cal Z}\eta}
            \bigg\}G^-_{22},\\
\label{eq:F_eq2}
\eea
\ew
where we have taken into account the relations
\bea
\nonumber
&&G_{12}=G_{21},\quad G_{11}=G_{22},\\
\label{eq:relation}
&&G^-_{12}=G^-_{21},\quad G^-_{11}=G^-_{22}.
\eea

 The poles of the zero-order Green's functions $G$ have to be the same
as the poles found for the first-order ones $G^{(1)}$.
 This can be seen directly by comparing the structure of the systems of equations for
the corresponding quantities: the system in Eq.~(\ref{eq:Z_eq}) for the zero-order
functions is identical with the systems in Eqs.~(\ref{eq:F_eq},\ref{eq:F_eq2}) for
the first-order ones, except for the free terms.
 The free terms in the first-order systems are determined by the zero-order Green's
functions, thus, the first-order quantities $G^{(1)}$ can be written down in terms
of the solution for the zero-order system of Eq.~(\ref{eq:Z_eq}), and the as yet unknown
quantities ${\rm v}_1$ and ${\rm v}_2$.

 To calculate ${\rm v}_{1,2}$ we use a relation connecting ${\rm v}$ and
the Green's functions $G^{(1)}(\bk,0^-)$.
From the definitions in Eq.~(\ref{eq:GGG}) and the expansion in
Eq.~(\ref{eq:explanation2}) we have
\begin{eqnarray}
G^f_{ii}(0^-)=\frac 12 - \langle\sigma^z_i\rangle^f=\frac 12 - \eta - f{\rm v}_i,
\end{eqnarray}
while the expansion in Eq.~(\ref{eq:explanation1}) leads to
\be
G^f_{ii}(0^-)=G_{ii}(0^-)+fG^{(1)}_{ii}(0^-)
= \frac 12 - \eta + fG^{(1)}_{ii}(0^-).
\ee
Thus, we can write down $-{\rm v}_i=G^{(1)}_{ii}(0^-)$, and after Fourier summation
one obtains
\bea
\label{eq:S1}
-{\rm v}_1&=&\frac 2N\sum_{\bk}G^{(1)}_{11}(\bk,0^-),\\
\label{eq:S2}
-{\rm v}_2&=&\frac 2N\sum_{\bk}G^{(1)}_{22}(\bk,0^-).
\eea
The solution of the system in Eq.~(\ref{eq:F_eq}) gives us the first-order Green's function
$G^{(1)}_{22}(\bk,\omega_m)$ and therefore ${\rm v}_2$.
Similarly, to find ${\rm v}_1$ we use Eq.~(\ref{eq:F_eq2}).

The solution of the system of equations in Eq.~(\ref{eq:Z_eq}) for the
zeroth-order Green's functions turns out to be
\bw
\bea
\nonumber
   \hspace*{-.7cm}
   G_{12}(\bk,\omega_n)=&&\!\!\!\!\!-\frac {\eta}2
   \left\{
   \left(1+\frac{J_2/2{+}A_{\bk}}{\omega_1(\bk)}\right)\frac 1{{\rm i}\omega_n{-}\varepsilon_1(\bk)}
  +\left(1-\frac{J_2/2{+}A_{\bk}}{\omega_1(\bk)}\right)\frac 1{{\rm i}\omega_n{+}\varepsilon_1(\bk)}
   \right.\\
\nonumber
   \hspace*{-.7cm}
 && \left.\quad
  -\left(1+\frac{J_2/2{-}A_{\bk}}{\omega_2(\bk)}\right)\frac 1{{\rm i}\omega_n{-}\varepsilon_2(\bk)}
  -\left(1-\frac{J_2/2{-}A_{\bk}}{\omega_2(\bk)}\right)\frac 1{{\rm i}\omega_n{+}\varepsilon_2(\bk)}
   \right\},\\
\nonumber
   \hspace*{-.7cm}
   G_{22}(\bk,\omega_n)=&&\!\!\!\!\!-\frac {\eta}2
   \left\{
   \left(1+\frac{J_2/2{+}A_{\bk}}{\omega_1(\bk)}\right)\frac 1{{\rm i}\omega_n{-}\varepsilon_1(\bk)}
  +\left(1-\frac{J_2/2{+}A_{\bk}}{\omega_1(\bk)}\right)\frac 1{{\rm i}\omega_n{+}\varepsilon_1(\bk)}
   \right.\\
\nonumber
   \hspace*{-.7cm}
 && \left.\quad
  +\left(1+\frac{J_2/2{-}A_{\bk}}{\omega_2(\bk)}\right)\frac 1{{\rm i}\omega_n{-}\varepsilon_2(\bk)}
  +\left(1-\frac{J_2/2{-}A_{\bk}}{\omega_2(\bk)}\right)\frac 1{{\rm i}\omega_n{+}\varepsilon_2(\bk)}
   \right\},\\
\nonumber
   \hspace*{-.7cm}
  G^-_{12}(\bk,\omega_n)=&&\!\!\!\!\!-\frac {\eta}2 B^*_{\bk}
  \left\{\!
  \frac 1{\omega_1(\bk)}\left(\frac 1{{\rm i}\omega_n{-}\varepsilon_1(\bk)}
                             -\frac 1{{\rm i}\omega_n{+}\varepsilon_1(\bk)}\right)
 +\frac 1{\omega_2(\bk)}\left(\frac 1{{\rm i}\omega_n{-}\varepsilon_2(\bk)}
                             -\frac 1{{\rm i}\omega_n{+}\varepsilon_2(\bk)}\right)
      \!\right\},\\
\nonumber
    \hspace*{-.7cm}
  G^-_{22}(\bk,\omega_n)=&&\!\!\!\!\!-\frac {\eta}2 B^*_{\bk}
  \left\{\!
  \frac 1{\omega_1(\bk)}\left(\frac 1{{\rm i}\omega_n{-}\varepsilon_1(\bk)}
                             -\frac 1{{\rm i}\omega_n{+}\varepsilon_1(\bk)}\right)
 -\frac 1{\omega_2(\bk)}\left(\frac 1{{\rm i}\omega_n{-}\varepsilon_2(\bk)}
                             -\frac 1{{\rm i}\omega_n{+}\varepsilon_2(\bk)}\right)
      \!\right\},\\
\label{eq:G_sol}
\eea
where the spectra for the out-of-plane $\varepsilon_1(\bk)$ and in-plane
$\varepsilon_2(\bk)$ modes\cite{Koshibae} are given by
\bea
\nonumber
   \hspace*{-.9cm}
&&\varepsilon_1(\bk)=2{\cal Z}\eta\omega_1(\bk)=2{\cal Z}\eta\sqrt{(J_2/2+A_{\bk})^2-|B_{\bk}|^2},\\
\nonumber
   \hspace*{-.9cm}
&&\varepsilon_2(\bk)=2{\cal Z}\eta\omega_2(\bk)=2{\cal Z}\eta\sqrt{(J_2/2-A_{\bk})^2-|B_{\bk}|^2}.\\
\label{eq:spectra}
\eea

After the substitution of the results in Eq.~(\ref{eq:G_sol})
into the system of equations in Eqs.~(\ref{eq:F_eq},\ref{eq:F_eq2}), and
then using the solutions for $G^{(1)}_{11}(\bk,\omega_m)$,
$G^{(1)}_{22}(\bk,\omega_m)$, the results for quantities
${\rm v}_1$ and ${\rm v}_2$ are found to be
\begin{eqnarray}
\label{eq:SS}
 {\rm v}_1-{\rm v}_2&=&\frac {\eta^2C_1}{1+4\eta^2J_2C_1},\\
 {\rm v}_1+{\rm v}_2&=&\frac {\eta^2C_2}{1-8\eta^2C_3},
\end{eqnarray}
where
\begin{eqnarray}
\nonumber
 C_1&=&\frac 2N\sum_{\bk}2\left\{
     \left(1+\frac{(J_2/2)^2{-}A_{\bk}^2{-}|B_{\bk}|^2}{\omega_1(\bk)\omega_2(\bk)}\right)
     \frac{n(\varepsilon_1){-}n(\varepsilon_2)}{\varepsilon_2(\bk){-}\varepsilon_1(\bk)}
     -\left(1-\frac{(J_2/2)^2{-}A_{\bk}^2{-}|B_{\bk}|^2}{\omega_1(\bk)\omega_2(\bk)}\right)
     \frac{n(\varepsilon_1){+}n(\varepsilon_2){+}1}{\varepsilon_1(\bk)+\varepsilon_2(\bk)}
     \right\},\\
\nonumber
 C_2&=&\frac 2N\sum_{\bk}\left\{\!
      \frac{(J_2/2{+}A_{\bk})^2\beta/2}
           {\omega^2_1(\bk)\sinh^2\frac{\beta\varepsilon_1}2}
     +\frac{|B_{\bk}|^2[2n(\varepsilon_1){+}1]}
           {\omega^2_1(\bk)\varepsilon_1(\bk)}
     +\frac{(J_2/2{-}A_{\bk})^2\beta/2}
           {\omega^2_2(\bk)\sinh^2\frac{\beta\varepsilon_2}2}
     +\frac{|B_{\bk}|^2[2n(\varepsilon_2){+}1]}
           {\omega^2_2(\bk)\varepsilon_2(\bk)}
      \!\right\},\\
\nonumber
 C_3&=&\frac 2N\sum_{\bk}\left\{\!
      \frac{(J_2/2{+}A_{\bk})\beta/2}{\sinh^2\frac{\beta\varepsilon_1}2}
     +\frac{(J_2/2{-}A_{\bk})\beta/2}{\sinh^2\frac{\beta\varepsilon_2}2}
     \!\right\},\quad\mbox{here}\;\;
     n(\varepsilon_{1,2})=[{\rm exp}(\beta\varepsilon_{1,2}(\bk))-1]^{-1}.\\
\label{eq:C1}
\end{eqnarray}
\ew

Now let us find the quantities which determine a linear response to a
magnetic field applied to the one of sublattice -- see
Eqs.~(\ref{eq:iterpret1},\ref{eq:iterpret2}). The longitudinal $z$ components of
the susceptibility in the characteristic representation are given by
\be
\label{eq:chi_z}
\chi^{\sigma^z\sigma^z}_{11}
=\frac{\partial \langle\sigma^z_1\rangle^f}{\partial f}\Big|_{f=0}={\rm v}_1,\;\;\;
\chi^{\sigma^z\sigma^z}_{12}
=\frac{\partial \langle\sigma^z_2\rangle^f}{\partial f}\Big|_{f=0}={\rm v}_2,
\ee
where the expansion of Eq.~(\ref{eq:explanation2}) was used.
The transverse $x$ and $y$ components of the susceptibility tensor are determined
in the terms of Green's functions as
\begin{eqnarray}
\nonumber
\hspace*{-.7cm}
 \chi^{\sigma^\alpha\sigma^{\alpha'}}_{11}&=&
 \frac 2N\sum_{i,i'}\int^{\beta}_0\!\!\langle T_{\tau}\sigma_i^{\alpha}(\tau)
 \sigma_{i'}^{\alpha'}(0)\rangle{\rm d}\tau,\\
\nonumber
\hspace*{-.7cm}
 \chi^{\sigma^{\alpha}\sigma^{\alpha'}}_{12}&=&
 \frac 2N\sum_{i,j}\int^{\beta}_0\!\!\langle T_{\tau}\sigma_i^{\alpha}(\tau)
 \sigma_{j}^{\alpha'}(0)\rangle{\rm d}\tau,\\
\label{eq:def}
\end{eqnarray}
where $\alpha=x,y$.
By substituting the solutions in Eq.~(\ref{eq:G_sol}) into the
definition in Eq.~(\ref{eq:def}) for the transverse components of
susceptibility, we easily obtain {\emph{exactly the same result}}
that we have already found within our MFA calculations --
that is, Eq.~(\ref{eq:all_chi_sigma}).

Then, using Eqs.~(\ref{eq:CRtoINx})-(\ref{eq:CRtoINz}) the components of the susceptibility in the
initial coordinate system of Eq.~(\ref{eq:H_DM}) are found to be
\begin{eqnarray}
\label{eq:SxSx}
\hspace*{-.7cm}
\chi^{x}&=&\frac 14 \frac 1{J_1+J_2},\\
\label{eq:SySy}
\hspace*{-.7cm}
\chi^{y}&=&\frac 14 \frac {\sin^2(\theta)}{J_2-J_3}
           +\cos^2(\theta)[{\rm v}_1-{\rm v}_2],\\
\label{eq:SzSz}
\hspace*{-.7cm}
\chi^{z}&=&\frac 14 \frac {\cos^2(\theta)}{J_2+J_3}
           +\sin^2(\theta)[{\rm v}_1+{\rm v}_2].
\end{eqnarray}

For completeness, we mention that
we have also performed the theoretical investigation of this model (\ref{eq:H_DM})
within spin-wave (SW) theory, and the final result for the components of static
susceptibility turns out to be
\begin{eqnarray}
\label{eq:SxSx_SW}
\hspace*{-.7cm}
\chi^{x~SW}&=&\frac 14 \frac 1{J_1+J_2},\\
\label{eq:SySy_SW}
\hspace*{-.7cm}
\chi^{y~SW}&=&\frac 14 \frac {\sin^2(\theta)}{J_2-J_3}
      +\cos^2(\theta) S^2C_1\big|_{\eta\to S},\\
\label{eq:SzSz_SW}
\hspace*{-.7cm}
\chi^{z~SW}&=&\frac 14 \frac {\cos^2(\theta)}{J_2+J_3}
           +\sin^2(\theta) S^2C_2\big|_{\eta\to S}.
\end{eqnarray}
It can be noted that the difference in the results within the RPA,
Eqs.~(\ref{eq:SxSx})-(\ref{eq:SzSz}), and spin-wave theory,
Eqs.~(\ref{eq:SxSx_SW})-(\ref{eq:SzSz_SW}),
came from the calculation of the components of the susceptibility in the direction
of the sublattice magnetization
(that is $\chi^{\sigma^z\sigma^z}_{11}$ and $\chi^{\sigma^z\sigma^z}_{12}$).
The spin-wave theory gives unity in the denominator of the expressions for
$\chi^{\sigma^z\sigma^z}_{11}$ and $\chi^{\sigma^z\sigma^z}_{12}$ in Eq.~(\ref{eq:SS}),
and $S=1/2$ instead of the order parameter $\eta$ everywhere in the numerator.
The similar situation takes place for antiferromagnetic Heisenberg model within
the RPA scheme\cite{Liu} and spin-wave theory.\cite{Liu2}

We also mention that the transverse components of the susceptibility
in the characteristic representation (\ref{eq:all_chi_sigma}) are equal
within the MFA, RPA, and SW theories.

\subsection{Related Thermodynamic quantities}

In order for the above RPA theory to be complete, we need to
determine the behaviour of the order parameter and the transition
temperature.

The above expressions for the components of susceptibility Eqs.~(\ref{eq:SySy},\ref{eq:SzSz}),
and for the elementary excitations (spin waves) given by
Eq.~(\ref{eq:spectra}), include the as-yet-unknown value of the order parameter $\eta$.
From the definition on the Green's functions we can obtain
\begin{eqnarray}
\label{eq:temp}
&&G_{nn}(\tau=0^-)=\langle \sigma^-_n\sigma^+_n\rangle=\frac 12 - \eta,\quad\\
\nonumber
\mbox{where}\quad
&&G_{nn}(0^-)=\frac 2N\sum_{\bk}G_{22}(\bk,0^-).
\end{eqnarray}
Substituting $G_{22}(\bk,\omega)$ from Eq.~(\ref{eq:G_sol}), and performing the
summation on the Matsubara frequencies, the equation on the order parameter
turns out to be
\begin{eqnarray}
\nonumber
\hspace*{-.9cm}
  \frac 1{\eta}=\frac 2N\sum_{\bk}
 \bigg\{&&\!\!\!\!\!\frac{J_2/2{+}A_{\bk}}{\omega_1(\bk)}
                    [2n(\varepsilon_1){+}1]\\
\label{eq:sigma^z}
\hspace*{-.9cm}
           && +\frac{J_2/2{-}A_{\bk}}{\omega_2(\bk)}
             [2n(\varepsilon_2){+}1] \bigg\}.
\end{eqnarray}
Since order parameter (\ref{eq:sigma^z}) (sublattice magnetization)
is temperature dependent, it follows that the spectrum of elementary
excitations (Eq.~(\ref{eq:spectra})) is also temperature dependent.

The N\'eel temperature at which $\eta$ vanishes within the adopted RPA approximation
is determined by
\be
\label{eq:T_n}
T_{\rm N}{=}\left\{\frac 14 \frac 2N\sum_{\bk}
           \left(\frac{J_2/2{+}A_{\bk}}{\omega^2_1(\bk)}
            +\frac{J_2/2{-}A_{\bk}}{\omega^2_2(\bk)}
           \right)\right\}^{-1}\!\!\!.
\ee
By putting $\eta\to 0$ we can find that $z$-component of susceptibility
$\chi^{z}$ in Eq.~(\ref{eq:SzSz})
\be
\chi^{z}\Big|_{\eta\to 0}=\frac 14 \frac {\cos^2(\theta)}{J_2+J_3}
                +\sin^2(\theta)\frac{TC_2|_{\eta\to 0}}{1{-}TT_{\rm N}},
\ee
diverges at the N\'eel temperature, whereas other components of
susceptibility remain finite as the N\'eel point is approached from below.

\subsection{Susceptibility in the paramagnetic case}

When the temperature of the system is above the N\'eel temperature, $T_N$,
there still exists short-range magnetic order.
To model such an order\cite{Liu} we introduce a fictitious field $h$ pointing in
the direction of the sublattice magnetization, that is the $z$ direction
in the characteristic representation.
To this end, the Hamiltonian
\begin{eqnarray}
\label{eq:H_para}
 H_h = H_{\rm CR} - h \sum_i\sigma^z_i - h \sum_j\sigma^z_j
\end{eqnarray}
is used, and the limit $h\to 0$ is taken after the calculation is
carried out.
To obtain the susceptibility above the N\'eel temperature, it is convenient
to introduce an order parameter defined by
\begin{eqnarray}
\label{eq:y_def}
y = \lim_{h\to 0}(2{\cal Z}\eta/h).
\end{eqnarray}

The calculations for the model are very similar to the ones above presented.
It is easy to show that paramagnetic version of the equation on the order
parameter in Eq.~(\ref{eq:sigma^z}) leads to
\begin{eqnarray}
\nonumber
\hspace*{-.9cm}
\frac 1y = \frac 2N\sum_{\bk}\frac {1}{{\cal Z}\beta}\bigg\{
           &&\!\!\!\!\!
           \frac{1+y(J_2/2{+}A_{\bk})}{(1+y(J_2/2{+}A_{\bk}))^2-y^2|B_{\bk}|^2}\\
\label{eq:eq_y}
\hspace*{-.9cm}
           &&\hspace*{-.5cm}
           +\frac{1+y(J_2/2{-}A_{\bk})}{(1+y(J_2/2{-}A_{\bk}))^2-y^2|B_{\bk}|^2}
           \bigg\}.
\end{eqnarray}
The quantity $y$ approaches to infinity as the temperature is lowered to
$T_{\rm N}$.
Indeed, putting $y\to\infty$ in Eq. (\ref{eq:eq_y}) we find the temperature
at which $y$  diverges, which is nothing but N\'eel temperature.

By a procedure similar to the above presented (that is, the RPA scheme below $T_N$)
the different components of the magnetic susceptibility in the paramagnetic
phase are found to be
\bea
\label{eq:SxSx_ab}
\chi^{x}&=&\frac 14 \frac 1{J_1+J_2+2/y},\\
\nonumber
\chi^{y}&=&\frac 14 \frac {\sin^2(\theta)}{J_2-J_3+2/y}
            +\cos^2(\theta)\frac{y^2 D_1}{1+8y(1{+}yJ_2/2)D_1},\\
\label{eq:SySy_ab}\\
\label{eq:SzSz_ab}
\chi^{z}&=&\frac 14 \frac {\cos^2(\theta)}{J_2+J_3+2/y}
           +\sin^2(\theta)\frac{y^2 D_2}{1-8y^2D_3},
\end{eqnarray}
where
\bw
\begin{eqnarray}
\nonumber
\hspace*{-.7cm}
&&D_1=\frac 1{2{\cal Z}^2\beta}\frac 2N\sum_{\bk}2\frac{(1{+}y(J_2/2{+}A_{\bk}))
                             (1{+}y(J_2/2{-}A_{\bk}))-y^2|B_{\bk}|^2 }
                {\{(1{+}y(J_2/2{+}A_{\bk}))^2-y^2|B_{\bk}|^2\}
                 \{(1{+}y(J_2/2{-}A_{\bk}))^2-y^2|B_{\bk}|^2\}},
\eea
\bea
\nonumber
\hspace*{-.7cm}
&&D_2=\frac 1{2{\cal Z}^2\beta}\frac 2N\sum_{\bk}\left[
                     \frac{(1{+}y(J_2/2{+}A_{\bk}))^2+y^2|B_{\bk}|^2}
                  {\{(1{+}y(J_2/2{+}A_{\bk}))^2-y^2|B_{\bk}|^2\}^2}+
             \frac{(1{+}y(J_2/2{-}A_{\bk}))^2+y^2|B_{\bk}|^2}
                  {\{(1{+}y(J_2/2{-}A_{\bk}))^2-y^2|B_{\bk}|^2\}^2}
              \right],\\
\nonumber
\hspace*{-.7cm}
&&D_3=\frac 1{2{\cal Z}^2\beta}\frac 2N\sum_{\bk}\left[
                      \frac{1{+}y(J_2/2{+}A_{\bk})}
              {(1{+}y(J_2/2{+}A_{\bk}))^2-y^2|B_{\bk}|^2}+
                      \frac{1{+}y(J_2/2{-}A_{\bk})}
              {(1{+}y(J_2/2{-}A_{\bk}))^2-y^2|B_{\bk}|^2}
              \right].\\
\end{eqnarray}
\ew
By putting $y\to\infty$ we obtain that the components of susceptibility
$\chi^{x}$ and $\chi^{y}$ are continuous at the
N\'eel point, whereas the
$z$-component of susceptibility diverges in the $y\to\infty$
limit at the N\'eel point, the latter result reflecting the presence
of the spontaneous canted ferromagnetic moment in the $z$ direction.

\subsection{Susceptibility in the $T=0$ limit}
\label{subsec:Teq0chis}

As we will present in the results discussion, the dimensionality of
the parameter space that seems to be relevant to the cuprates is large,
but there are only a few important values that determine the physical
properties of the system. Here we discuss two key experimentally
obtainable quantities, and their relation to the above theory.

It has been reported, using inelastic neutron scattering, that the out-of-plane
($\varepsilon_1$) and in-plane ($\varepsilon_2$) spin-wave gaps are $5.0$ and
$2.3$ meV, respectively, in the LTO phase of \LCO~crystal.\cite{Keimer}
Using these results let us predict the ratio of the components of susceptibility
$\chi^{y}/\chi^{x}$. The zone-centre (${\bk} = 0$) spin-wave gaps are given by
\bea
\nonumber
\varepsilon_1&=&{\cal Z}\eta\sqrt{(J_2+J_1)(J_2-J_3)},\\
\label{eq:var12}
\varepsilon_2&=&{\cal Z}\eta\sqrt{(J_2-J_1)(J_2+J_3)},
\eea
and they are real if $J_2<J_1,J_3$. So, from these relations we obtain
\be
\varepsilon_2<\varepsilon_1\quad\Leftrightarrow\quad J_1>J_3.
\ee
Also, in the $T=0$ limit the $y$ component of the susceptibility in
Eq.~(\ref{eq:MFA_SySy}) is given by
\begin{eqnarray}
\nonumber
\hspace*{-.6cm}
&&\chi^{y~ MFA}=\frac 14 \frac {\sin^2(\theta)}{J_2-J_3}=
          \frac 14\frac{J_2-J_1}{J^2_2-J^2_3},\\
\hspace*{-.6cm}
\mbox{then}&&
\frac{\chi^{y~MFA}}{\chi^{x~MFA}}=\frac{J^2_2-J^2_1}{J^2_2-J^2_3}.
\end{eqnarray}
Therefore, within the MFA
\be
\chi^{x~MFA}>\chi^{y~MFA}\quad\Leftrightarrow\quad J_1>J_3.
\ee
Thus, if $\varepsilon_2 < \varepsilon_1$ ($\varepsilon_2 > \varepsilon_1$),
in the limit of zero temperature the MFA predicts that $\chi^y < \chi^x$
($\chi^y > \chi^x$).

In the limit of the small anisotropy $d,\Gamma \ll J$ the components of the
susceptibility at $T=0$ within the MFA turn out to be
\begin{eqnarray}
\nonumber
&&\chi^{x~MFA}\approx\chi^{z~MFA}\approx\frac 1{8J}, \\
\label{chi:approximation}
&&\chi^{y~MFA}\approx\frac{d^2}{32J^2(\Gamma_1-\Gamma_3)},
\end{eqnarray}
while the expressions for the spin-wave gaps are
\begin{eqnarray}
\label{spectrum:approx}
\hspace*{-.7cm}
\varepsilon_1\approx{\cal Z}\eta\sqrt{2J(\Gamma_1-\Gamma_3)},\;
\varepsilon_2\approx{\cal Z}\eta d/\sqrt{2}.
\end{eqnarray}
We can see that components $\chi^{x,z}$ are almost independent of the anisotropy
parameters, while the $\chi^{y}$ component is very sensitive to the \emph{ratio}
between the anti-symmetric $d$ and symmetric $\Gamma_1-\Gamma_3$ parameters of
anisotropy. Then, the ratio between the components of the susceptibility is given by
\begin{eqnarray}
\label{ratio}
\frac{\chi^{x,z~MFA}}{\chi^{y~MFA}}\approx
\left(\frac{\varepsilon_1}{\varepsilon_2}\right)^2.
\end{eqnarray}

It can be noted that within the MFA scheme the different components of the
susceptibility, i.e. $\chi^x$, $\chi^y$ and $\chi^z$, are determined by the
contributions from the \emph{transverse components} of the susceptibility
in the characteristic representation. Indeed, as should be expected,
the longitudinal components of the susceptibility in the CR
(see Eq.~(\ref{eq:MFA_z2})) are equal to zero in the $T=0$ limit.
As showed earlier in this paper, in the characteristic representation
the RPA and SW theories lead to the same result for the transverse components
of the susceptibility as the MFA does. Since the longitudinal components in
the CR, given by the Eq.~(\ref{eq:SS}) within the RPA, and their simplified
expressions within the SW theory (see Eqs.~(\ref{eq:SySy_SW},\ref{eq:SzSz_SW}))
become negligibly small in the $T=0$ limit, we predict that RPA, SW and MFA
within the reasonable range of the model parameters ($d,\Gamma \ll J$) satisfy
the ratio of Eq.~(\ref{ratio}), and the different components of the susceptibility
at $T=0$ can be approximated by the Eq.~(\ref{chi:approximation}).

 We also note the analogy with the pure 3D Heisenberg model where, in the limit of zero temperature,
all approximations considered here give the same magnitude for the transverse components of the susceptibility,
and zero for the longitudinal one.\cite{Liu}

\section{Results of calculations}
\label{sec:Results}

In this section we present the results of a numerical investigation of the magnetic
properties of the system modelled by the Hamiltonian given by Eq.~(\ref{eq:H_DM})
based on the above presented analytical formulae. Specifically, we are interested in
the temperature dependencies of the various components of the susceptibility for
different values (specifically, ratios) of the model parameters. Further, we will
numerically demonstrate the correlation between the magnitudes of the two spin-wave
gaps in the excitation spectrum and the behaviour of the susceptibility components;
this relation was discussed analytically in the previous subsection. Also, and
most importantly, we will make clear the role played by quantum fluctuations by
comparing the results of the different approximation schemes.

In what follows, we will mainly examine one set of parameters that are suggested from
experimental measurements discussed in the previous subsection, and this will
allow us to ``zero in" on a parameter regime. However, since have developed
the theory for one plane and not a 3d solid, we do not necessarily expect this
set of parameters to be representative of a system like \LCO; instead, as
we discussed in the introduction to this paper, this approach will allow us to
determine if a one-plane approach is adequate, since, as we and others have
discussed, a true $T_c>0$ phase transition is possible for one plane and thus
could possibly be sufficient for this system.

In the present calculations we express all model parameters in terms of $J$.
Also, as will be made clear below, instead of using the set of parameters $\Gamma_1$,
$\Gamma_2$, and $\Gamma_3$, we deal with combination of parameters
$\Gamma_1-\Gamma_3$, $\Gamma_1+\Gamma_3$, and $\Gamma_2$.
The chosen magnitudes of the model parameters $d$ and $\overleftrightarrow{\Gamma}$
give the reported magnitude of gaps in the spectrum\cite{Keimer,gozar04}
\be
\label{eq:esps_expt}
\varepsilon_o=\varepsilon_1\approx 5 \mbox{ meV},\qquad
\varepsilon_i=\varepsilon_2\approx 2.3 \mbox{ meV},
\ee
at the temperature $T=T_N/3$ for the superexchange value $J=130$ meV.\cite{Keimer}
This leads to the parameters given by
\be
\label{eq:dGamma_fromgaps}
d/J = 0.02,~~~~~~~(\Gamma_1~-~\Gamma_3)/J~=~0.42\times 10^{-3},
\ee
where, as discussed below, we have set
\be
\label{eq:zero_Gammas}
(\Gamma_1~+~\Gamma_3)/J~=~0,~~~~~~~\Gamma_2~=~0.
\ee

\subsection{The AFM order parameter, spin-wave excitations, and T$_N$}

To examine the different analytical schemes used in the previous sections, we compare
the representative solutions of the order parameter, $\eta$, within the RPA method,
Eq.~(\ref{eq:sigma^z}), and within the MFA, Eq.~(\ref{eq:sigma_MFA}).
Most interestingly, we note the results shown in Fig.~{\ref{fig:eta_omega}(a) look
very similar to the corresponding ones for the pure 3D quantum Heisenberg
antiferromagnet within the RPA and MFA schemes.\cite{Tyablikov}

Since the order parameter $\eta$ is temperature dependent, it follows that within
the RPA scheme the spin-wave spectrum (see Eq.~(\ref{eq:spectra})) is also
temperature dependent. In Fig.~{\ref{fig:eta_omega}(b) we present the behaviour
of both modes in the excitation spectrum
at the long wavelength limit (${\bk}=0$)  (energy gaps) with respect to the
relative temperature ($T/T_N$); these results compare favourably with the
experimental measurements of the same quantity.\cite{Keimer}
\begin{figure}[h]
 \epsfxsize 0.34\textwidth\epsfbox{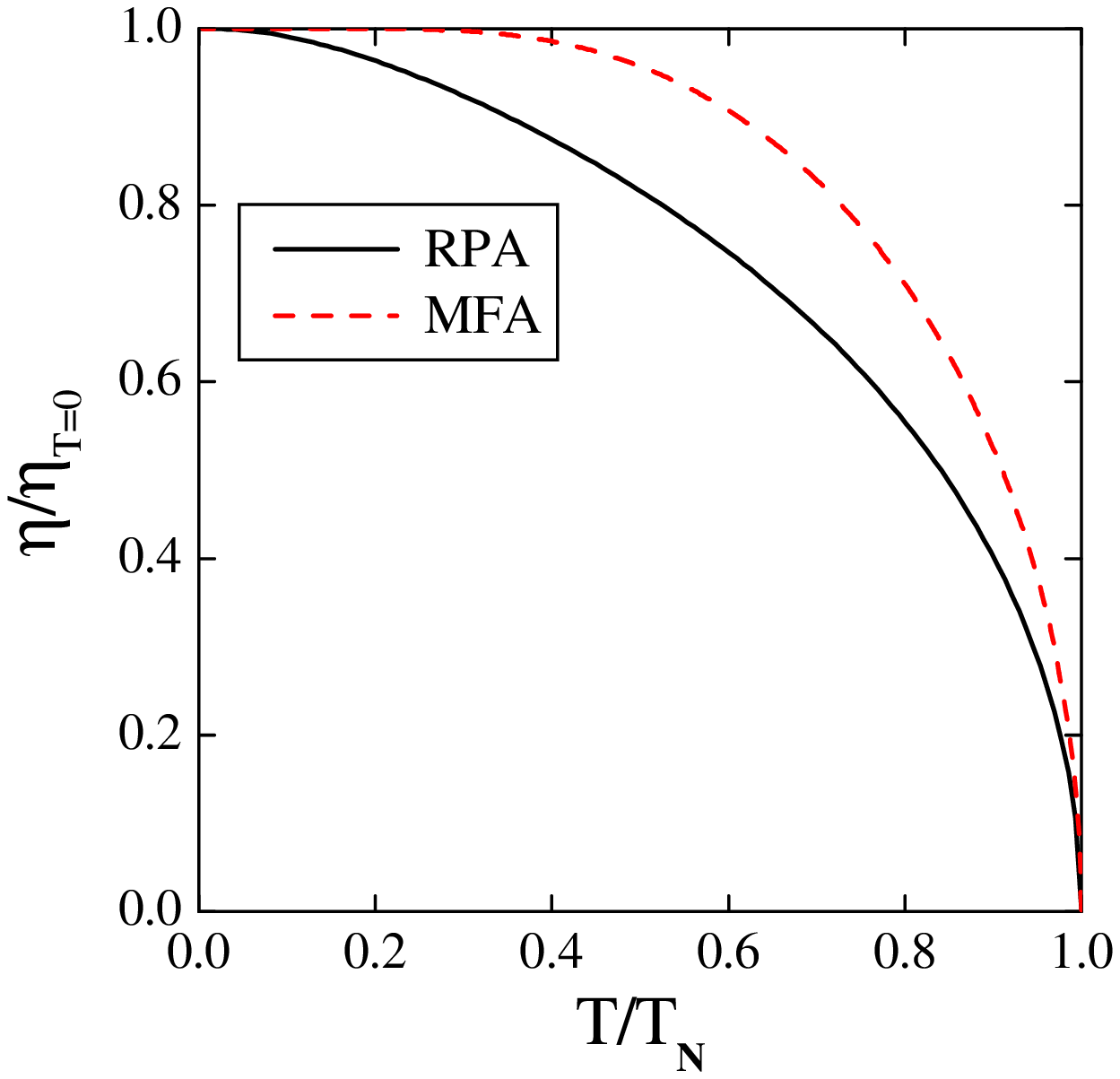}\;(a)\\
 \epsfxsize 0.34\textwidth\epsfbox{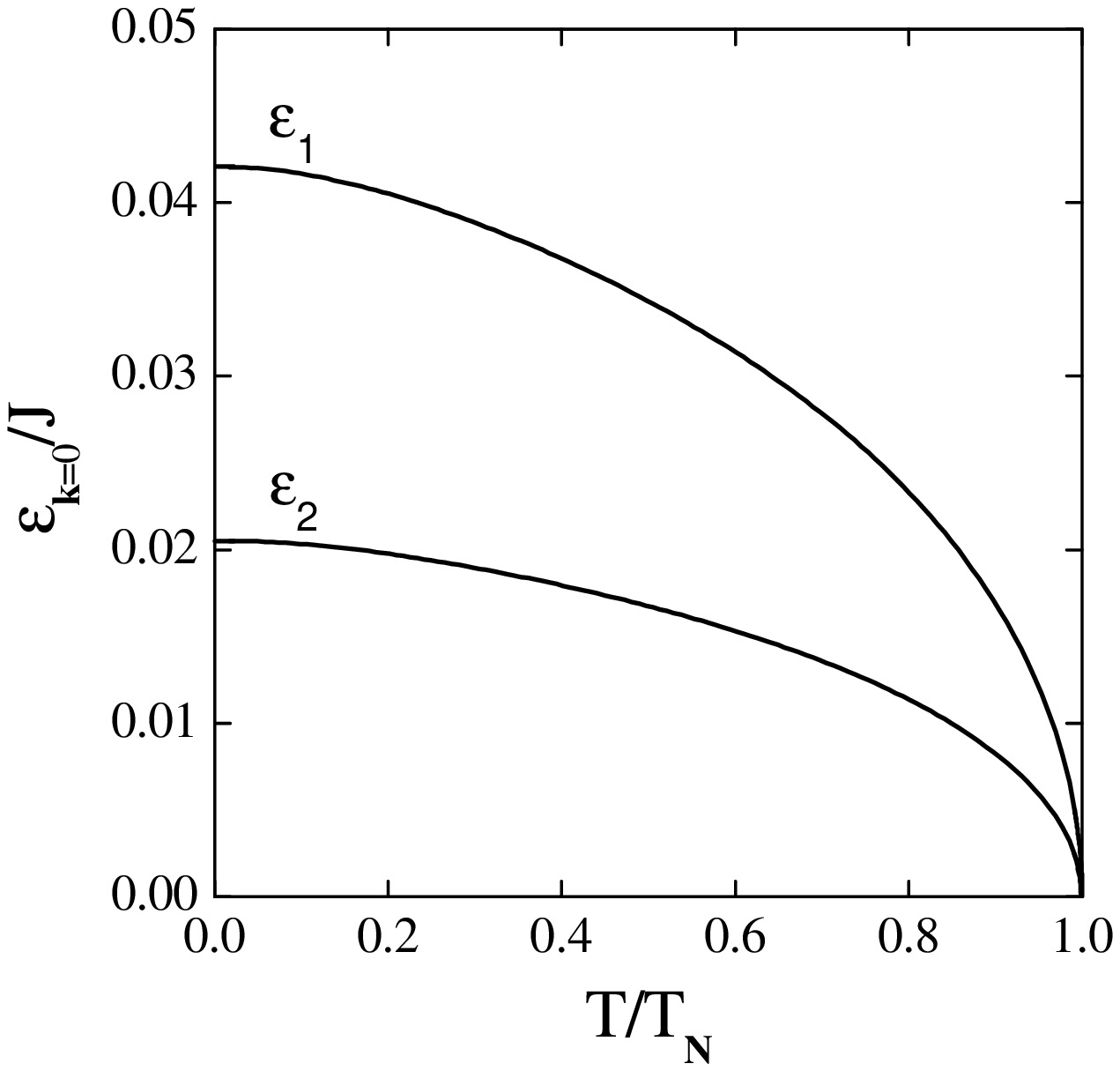}\;(b)
\caption{\label{fig:eta_omega} (Color online)
          The (a) order parameter \textit{vs.} $T/T_N$ within the RPA method (black solid line)
          and the MFA (red dashed line), and the (b) spin-wave gaps, in units of $J$, in the
          spectrum of elementary excitations  \textit{vs.} $T/T_N$ within
          the RPA method. In both of these figures we have used $d/J = 0.02$,
          $(\Gamma_1-\Gamma_3)/J = 0.42\times 10^{-3}$, $\Gamma_1+\Gamma_3 =0$, and $\Gamma_2 = 0$.}
\end{figure}

Now let us show that in contrast to the MFA approach,
where $T_{\rm N}=J_2\approx J$ is almost independent of the anisotropy, the N\'eel temperature
within the RPA analytical scheme is very sensitive to model parameters $d$ and $\Gamma_1-\Gamma_3$.
\begin{figure}[h]
 \epsfxsize 0.34\textwidth\epsfbox{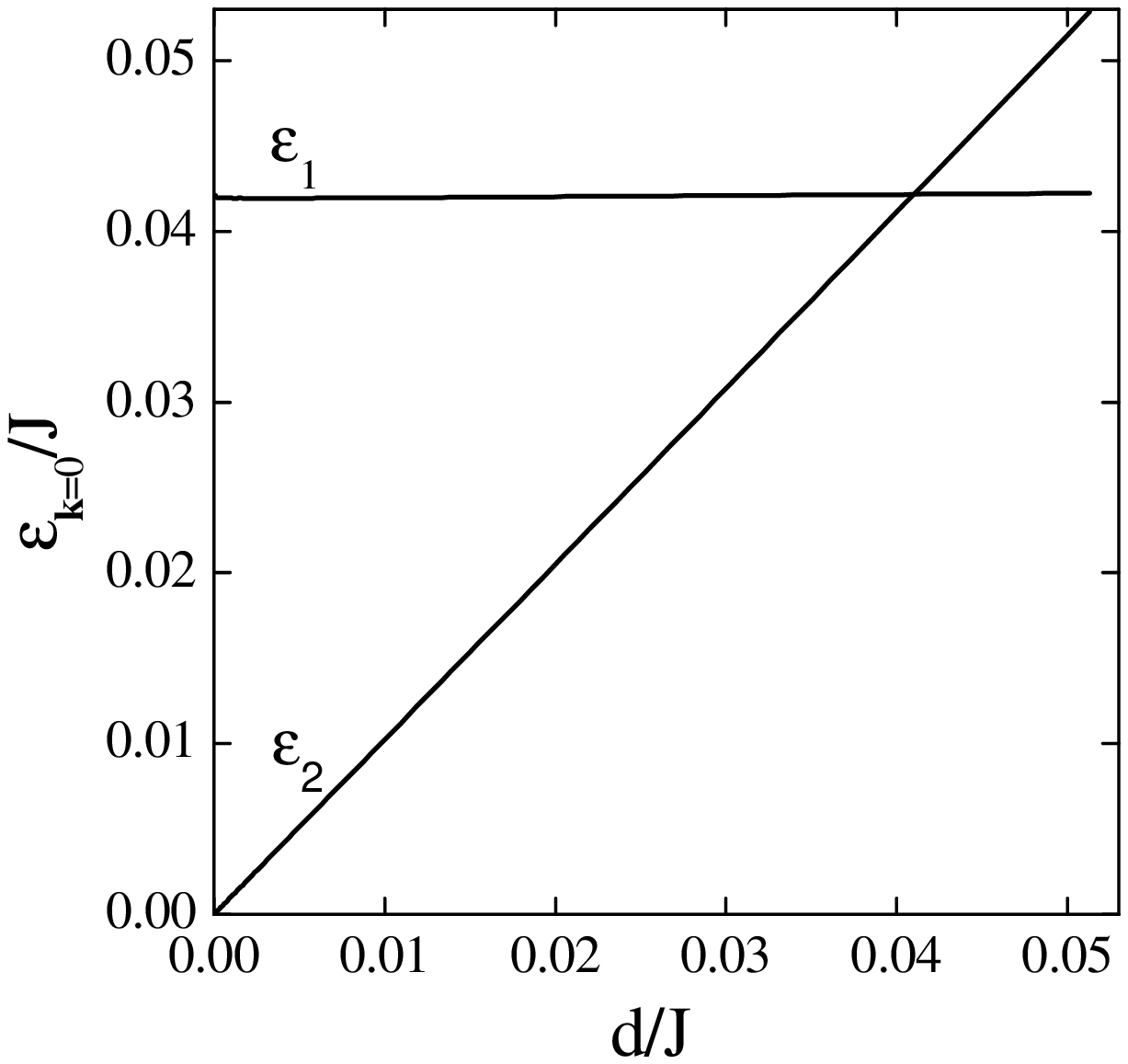}\;(a)\\
 \epsfxsize 0.34\textwidth\epsfbox{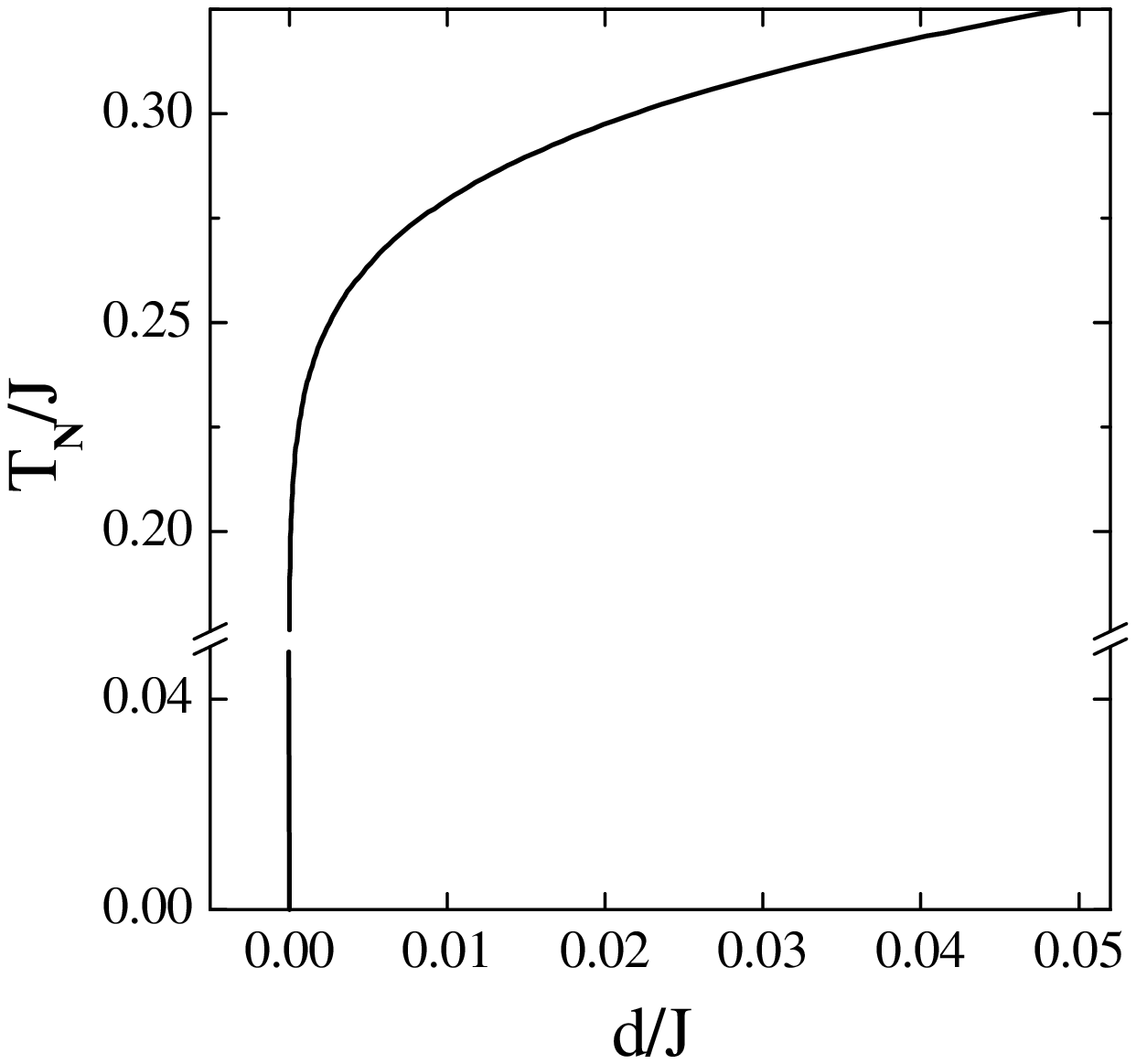}\;(b)
\caption{\label{fig:gaps_d} The (a) $T=0$ energy gaps, in units of $J$, \textit{vs.} the DM parameter
         $d/J$, as well as (b) the N\'eel temperature $T_N$, in units of $J$, \textit{vs} $d/J$.
         In both of these figures, $(\Gamma_1-\Gamma_3)/J = 0.42\times 10^{-3}$,
         $\Gamma_1+\Gamma_3 =0$, and $\Gamma_2 = 0$.}
\end{figure}
Figure~\ref{fig:gaps_d} shows the zero-temperature energy gaps and the N\'eel temperature
as functions of the DM antisymmetric exchange interaction $d/J$ within the RPA method.
As one can see, the energy gap $\varepsilon_1$ is almost independent of the $d/J$,
while $\varepsilon_2$ depends almost linearly on the DM interaction $d/J$, and
in fact goes to the zero in the limit $d/J \to 0$. As a result, when $d/J=0$ the
Goldstone mode appears in the spin-wave spectrum and thermal fluctuations
destroy the long-range ordering for any $T>0$. Consequently, the
N\'eel temperature drops to zero in case of $d=0$.

In the next two figures we present the dependencies on the model parameter $(\Gamma_1-\Gamma_3)/J$.
Now, the energy gap $\varepsilon_2$ is independent of the parameters of symmetric anisotropy
$\overleftrightarrow{\Gamma}$ and thus determined by the DM interaction $d/J$ alone, while the gap
$\varepsilon_1$ varies strongly with $(\Gamma_1-\Gamma_3)/J$.
\begin{figure}[h]
 \epsfxsize 0.34\textwidth\epsfbox{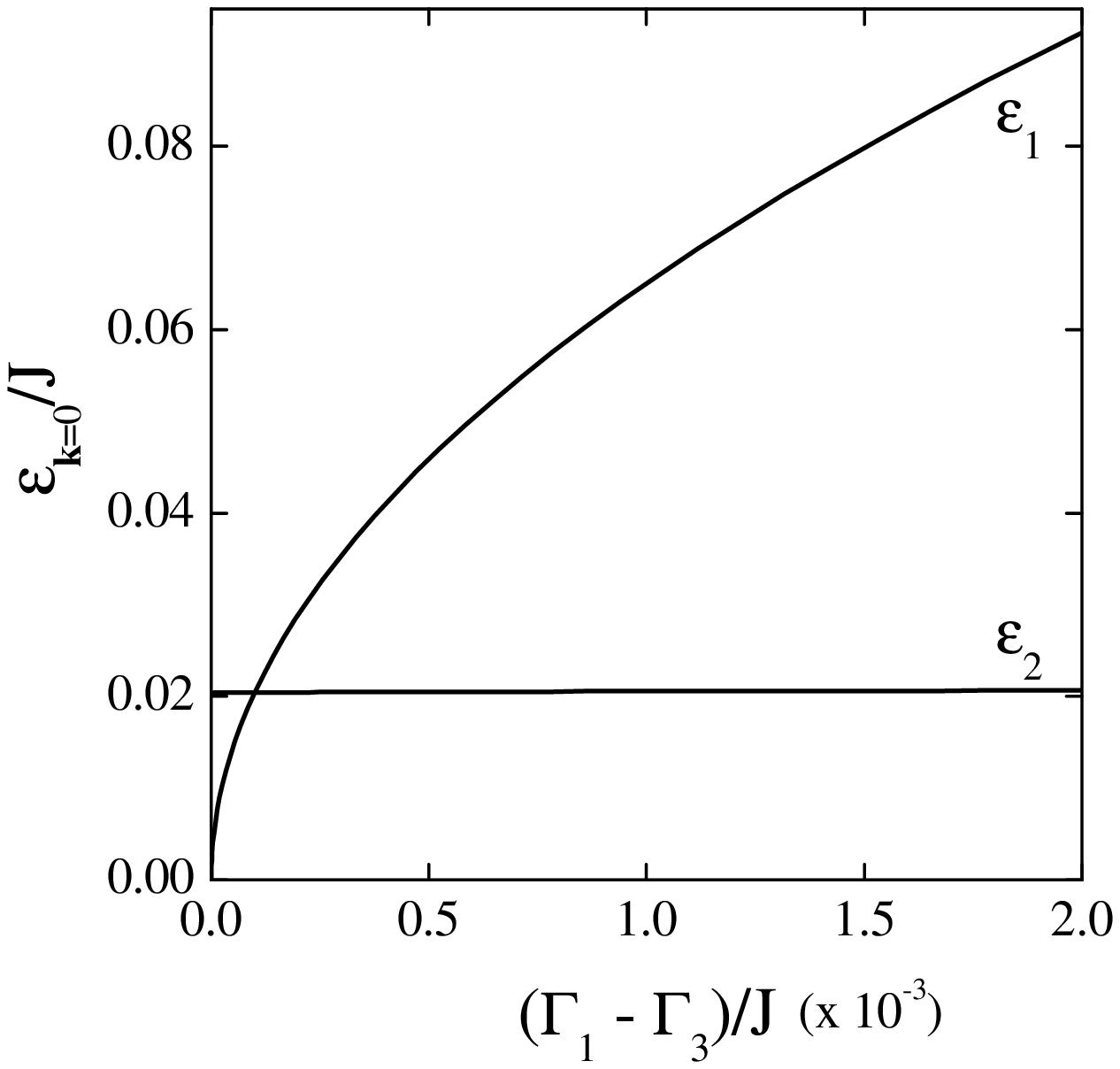}\;(a)\\
 \epsfxsize 0.34\textwidth\epsfbox{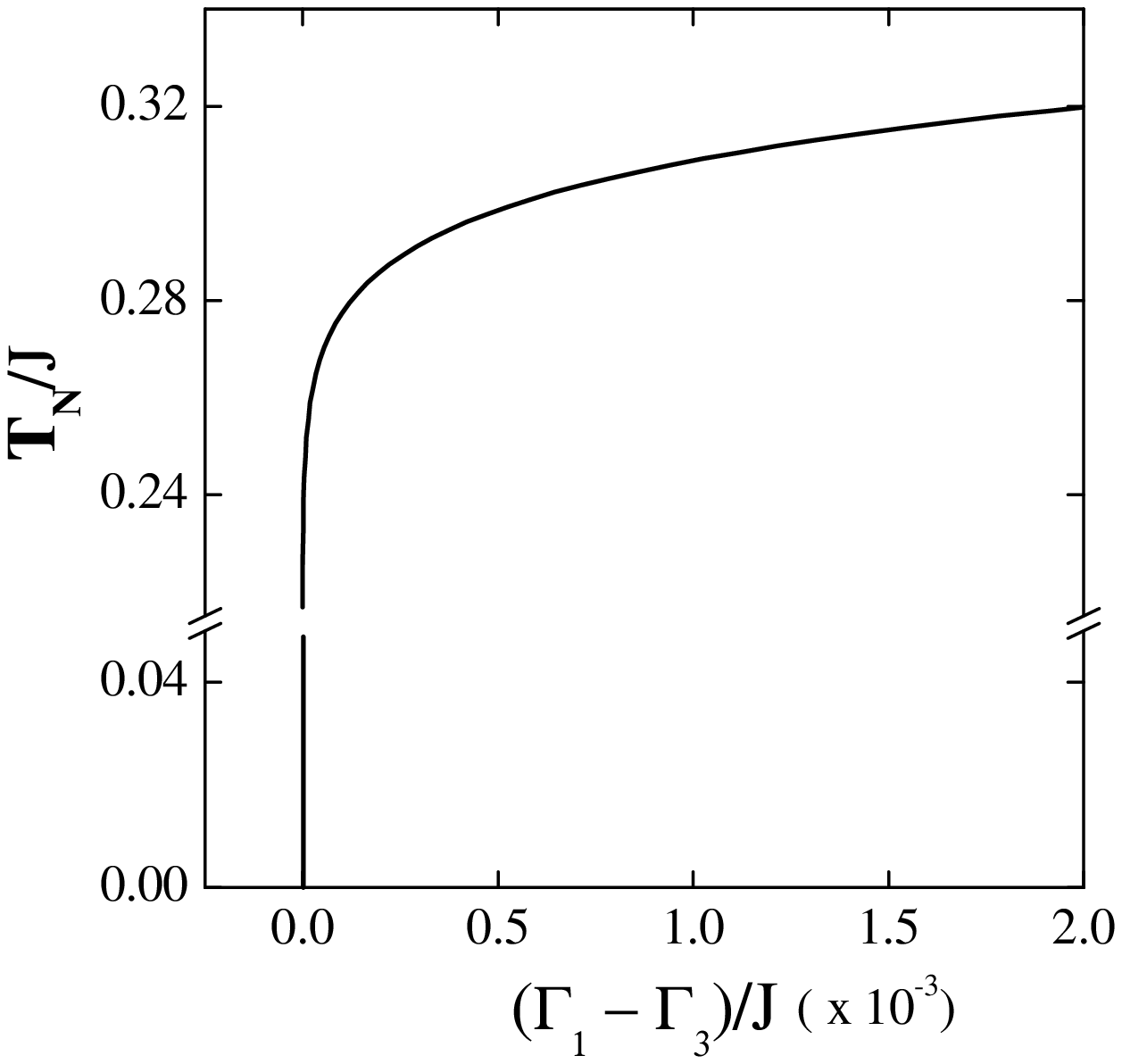}\;(b)
\caption{\label{fig:gaps_Gamma} The (a) energy gaps, in units of $J$, \textit{vs.} $(\Gamma_1-\Gamma_3)/J$,
          and (b) the N\'eel temperature, in units of $J$, as function of $(\Gamma_1-\Gamma_3)/J$.}
\end{figure}
As in the above case, in the limit of $\Gamma_1-\Gamma_3\to 0$ the mode
$\varepsilon_2$ in the spectrum becomes gapless and, therefore, the transition
temperature to the long-range ordered state would be suppressed to zero.

One can understand the above results for the zone-centre excitation spectrum
immediately from the
the expressions Eq.~(\ref{spectrum:approx}) in the limit of small anisotropy.
%\begin{eqnarray}
%\varepsilon_1\approx {\cal Z}\eta\sqrt{2J(\Gamma_1-\Gamma_3)},\qquad
%\varepsilon_2\approx {\cal Z}\eta d/\sqrt{2}.
%\end{eqnarray}
Our numerical results, shown in the previous figures, demonstrate
that these expressions are valid over a large range of parameter values.

\subsection{Parameters regimes}

We now summarize our numerical results with regards to the dependence of various thermodynamic
quantities on the material parameters appearing in the Hamiltonian.

Firstly, we find  that the N\'eel temperature is almost independent of the $(\Gamma_1+\Gamma_3)/J$ within
the reasonable range of the model parameters (see below).
\begin{figure}[h]
 \epsfxsize 0.34\textwidth\epsfbox{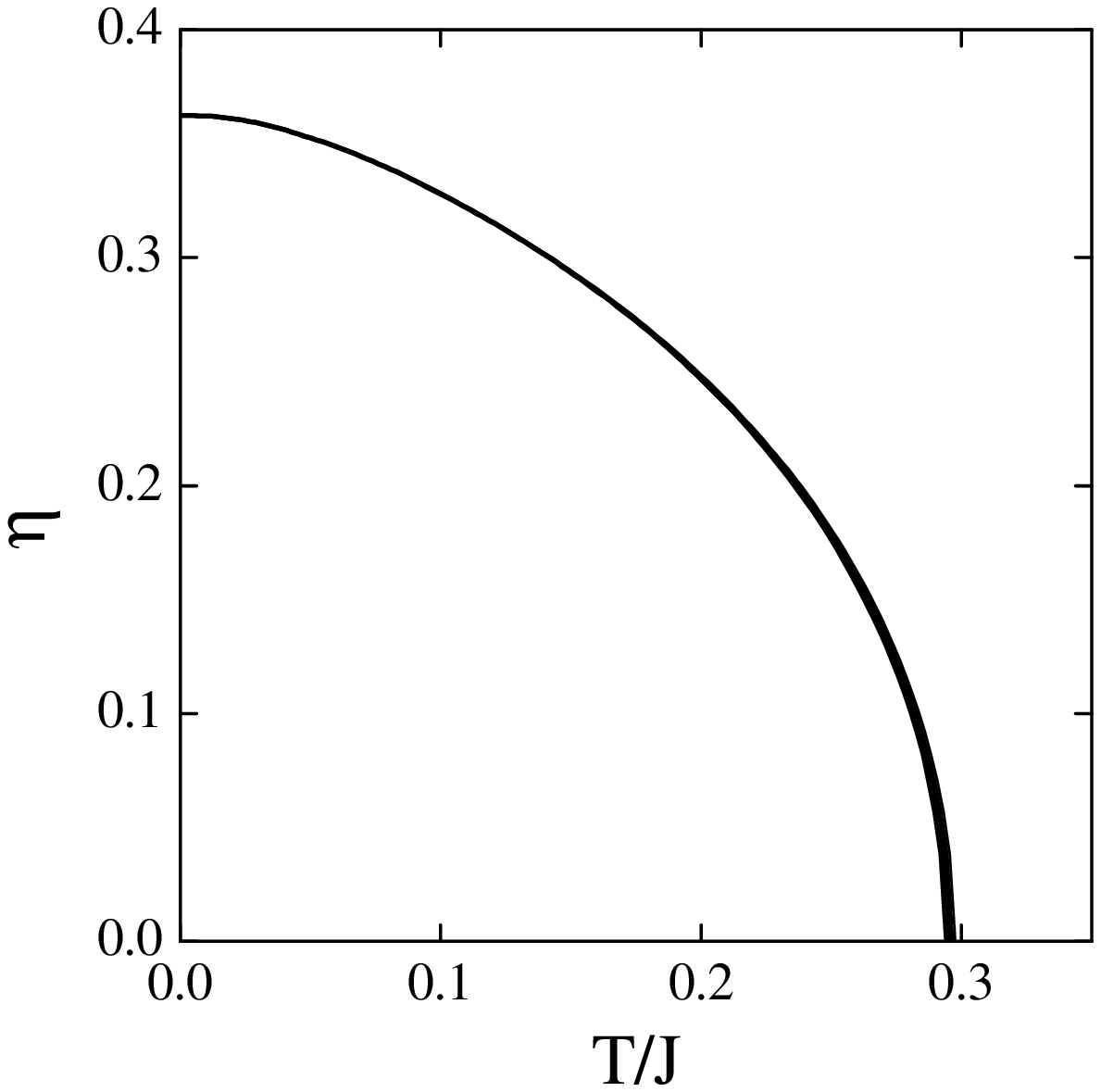}\;(a)\\
 \epsfxsize 0.34\textwidth\epsfbox{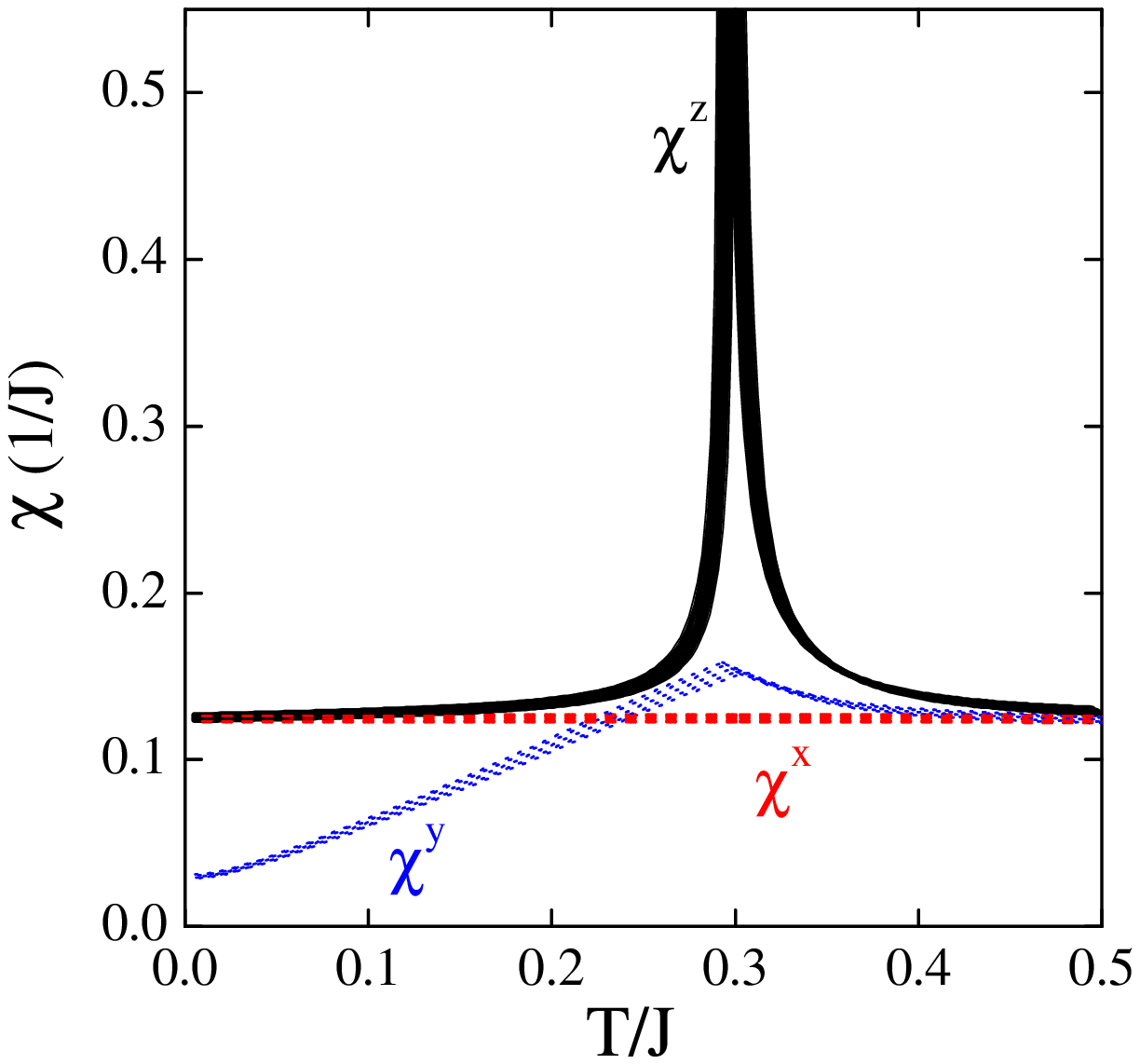}\;(b)
\caption{\label{fig:vary_params} (Color online) The (a) order parameter \textit{vs.} $T/J$ for the different values
           of $\Gamma_2/J$ and $(\Gamma_1+\Gamma_3)/J$, and (b) the susceptibility, in
           units of $1/J$,  \textit{vs.} $T/J$  for the
           different values of $\Gamma_2$ and $\Gamma_1+\Gamma_3$ discussed in the test. In these plots we have
           fixed $d/J = 0.02$, and $\Delta\Gamma/J \equiv (\Gamma_1-\Gamma_3)/J = 0.42\times 10^{-3}$).}
\end{figure}
In fact, in order to argue for the independence of the thermodynamic quantities central
to this study on certain material parameters that appear in the Hamiltonian, {\em viz.}
$\Gamma_2$ and $\Gamma_1+\Gamma_3$, in Fig.~\ref{fig:vary_params} we show two representative
plots for the order parameter and the susceptibility within the RPA scheme for the
constant values of the $d$ and $\Delta\Gamma=\Gamma_1-\Gamma_3$ (again in units of $J$).
That is, in each of the plots in Fig.~\ref{fig:vary_params} we have simultaneously plotted
ten data sets each with the different values of the $\Gamma_2$ and $\Gamma_1+\Gamma_3$,
where the parameter ratio $\Gamma_2$ has been varied from the value $-10^3\times\Delta\Gamma$
up to the $10^3\times\Delta\Gamma$, and $\Gamma_1+\Gamma_3$ from the value $-10^2\times\Delta\Gamma$ to the
$10^2\times\Delta\Gamma$ (all in units of $J$). As one can see, even for such a large range of the
parameters, one can hardly see the difference of the absolute values of the N\'eel temperature, order parameter
and susceptibility.

Thus, to study the magnetic properties of the system we can use only the DM interaction
$d$ and the combination $\Gamma_1-\Gamma_3$ of the symmetric tensor components as
two independent parameters, and so we conclude (similar to others\cite{gozar04,Peters})
that the system can be studied using $\Gamma_2=0$ and $\Gamma_1+\Gamma_3=0$.

In various limits, it can be shown that this result follows from the above presented
analytical work. The non-diagonal term $\Gamma_2$ of the symmetry anisotropy tensor is involved
in all expression through the combination in the $J_4$ (Eq.~(\ref{eq:JJJJ})). For the reasonable
anisotropy parameters (that is $\Gamma < d \ll J$) the spins are canted by a
very small angle
$\theta\sim d/J$,
and as a result we can neglect the term $\Gamma_2\sin\theta\sim d\Gamma_2/J$ with
respect to $d$, and hence we can ignore the quantity $\Gamma_2$ in all our formulae.
Similarly, $\Gamma_1+\Gamma_3$ is involved in the formulae through the combination
in the $J_2$, $J_3$, and canted angle $\theta$, where it appeared only as the
combination $J+\frac 12(\Gamma_1+\Gamma_3)$.
Thus, the parameter $\Gamma_1+\Gamma_3$ can be ignored with respect to the
superexchange interaction $J$ (see Eq.~(\ref{eq:JJJJ})).

Therefore, one can assert that the model Hamiltonian of Eq.~(\ref{eq:H_DM}) leads
to the same results as, for instance, the model described by the spin Hamiltonian
\be
  H=\sum_{\langle i,j\rangle}[J{\bf S}_i\cdot{\bf S}_j
    -\Delta\Gamma S^z_iS^z_j+{\bf D}_{ij}({\bf S}_i\times{\bf S}_j)],
\ee
where we define $\Delta\Gamma\equiv\Gamma_1-\Gamma_3$.

\subsection{Susceptibility}

Now let us consider the main focus of our paper, that being the behaviour of
the different components of static uniform magnetic susceptibility as a function
of temperature. Our results for $\chi^x$, for the parameters discussed
in the previous subsections, are shown in Fig.~\ref{fig:susc_x} (recall
our earlier result that the MFA and SW theories predict the same
$T$-independent value for this quantity).
The $x$ component of susceptibility below the N\'eel temperature
is the temperature independent and is equal to $\approx 1/(8J)$ within the MFA
(Eq.~(\ref{eq:MFA_SxSx})), the RPA scheme (Eq.~(\ref{eq:SxSx})), or spin-wave theory
(Eq.~(\ref{eq:SxSx_SW})). However, above the ordering temperature, the RPA and MFA
yield different results, with a weak $T$-dependence within the RPA, while a strong
Curie-like falloff is found within the MFA.
\begin{figure}[h]
 \epsfxsize 0.34\textwidth\epsfbox{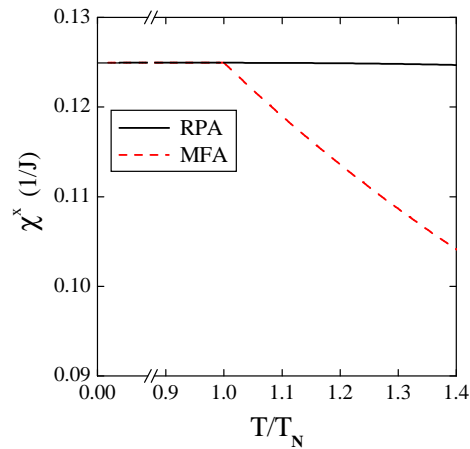}
\caption{\label{fig:susc_x} (Color online) The susceptibility $\chi^x$, in units
of $1/J$, within the RPA (black solid line) and MFA (red dashed line),
for the parameters values $d/J = 0.02$ and $\Delta\Gamma/J = 0.42\times 10^{-3}$.
Below $T_N$ these theories both predict the same constant value that is
independent of temperature.}
\end{figure}

As we will discuss in a future publication, this behaviour changes if one includes
4-spin ring exchange, or goes beyond the Tyablikov RPA decoupling scheme that we
employ in this paper. This is important since the experimental data of
Lavrov \textit{et al.},\cite{Ando} shows a small nonzero slope of
$\chi^x$ \textit{vs.} $T$. Indeed, a successful comparison with the small slope
seen below $T_N$ in experimental data\cite{Ando} necessarily requires that we
go beyond the treatment of spin-wave interactions and/or Hamiltonian that are
included in this paper. We emphasize that the necessity of going beyond the
Tyablikov RPA decoupling to obtain this slope is a manifestation of the presence
of strong quantum fluctuations, a theme that will be repeated in our discussion
in this and the next section of this paper.

Our results for the $y$ component of the susceptibility, $\chi^y$, are shown in
Fig.~\ref{fig:susc_y}.
\begin{figure}[h]
 \epsfxsize 0.34\textwidth\epsfbox{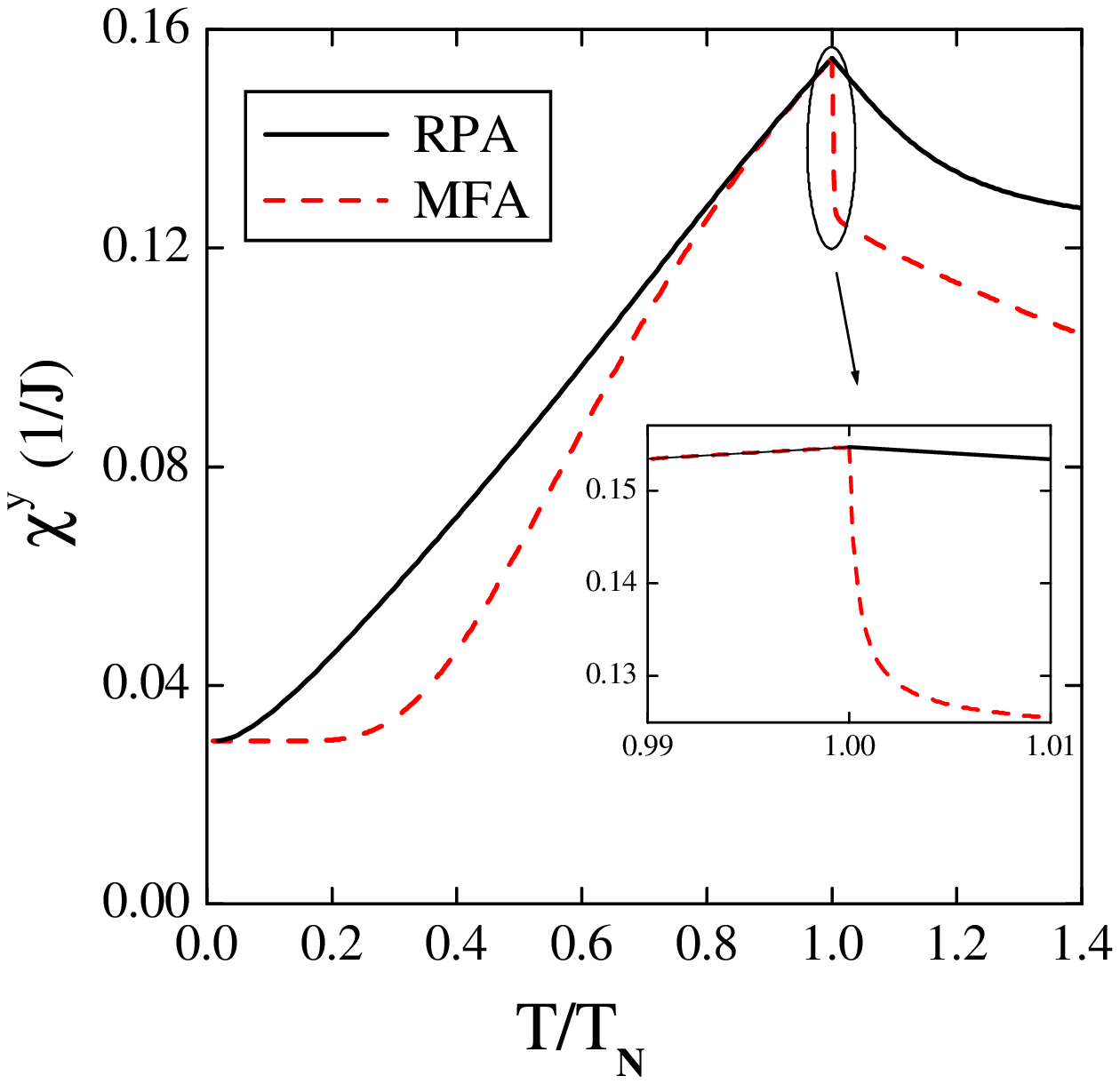}\;(a)\\
 \epsfxsize 0.34\textwidth\epsfbox{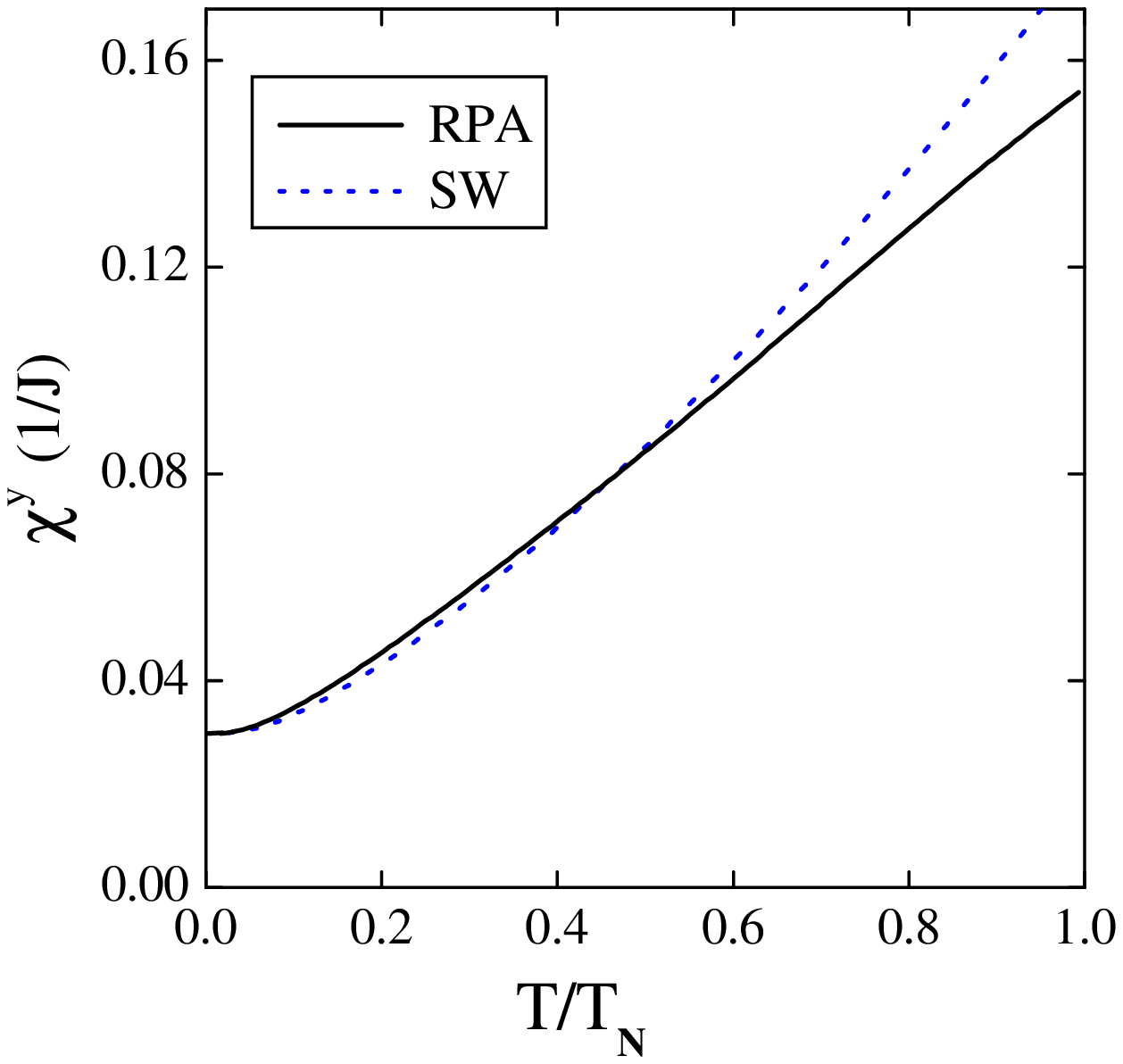}\;(b)
\caption{\label{fig:susc_y} (Color online)
The (a) susceptibility $\chi^y$ within the RPA
  (black solid line) and MFA (red dashed line), and (b) a comparison of the RPA
  (black solid line) and  spin-wave (SW) (blue dotted line) results below $T_N$.
   As in previous figures, we are using $d/J = 0.02$
   and $\Delta\Gamma/J = 0.42\times 10^{-3}$.}
\end{figure}
These plots show that below the ordering temperature the RPA scheme
leads to the good agreement with the MFA scheme near the $T_N$ ($0.8T_N<T<T_N$),
and good agreement with the SW theory at low-$T$ (that is, for $T<T_N/2$).
Above the N\'eel temperature, the  RPA and MFA theories lead to
very different results. The MFA method gives an abrupt decrease of $\chi^y$ to
a value that is close to that of the purely transverse component
$\chi^x\approx 1/(8J)$ (see inset of this figure),
while the RPA leads to a much more gradual decrease of the value of $\chi^y$
with the temperature.

The $z$ component of the susceptibility, $\chi^z$, is shown in Fig.~\ref{fig:susc_z}.
\begin{figure}[h]
 \epsfxsize 0.34\textwidth\epsfbox{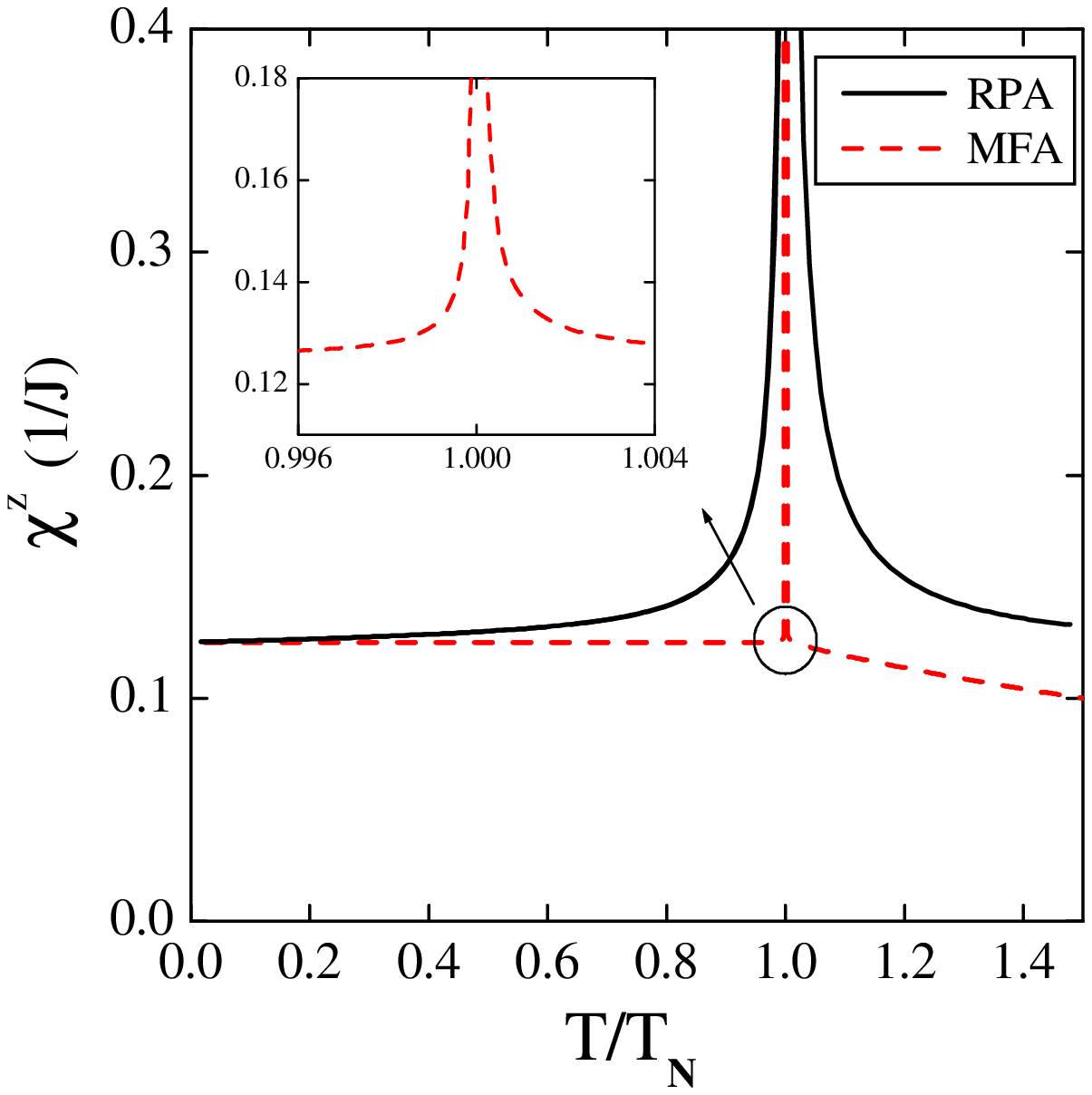}\;(a)\\
 \epsfxsize 0.34\textwidth\epsfbox{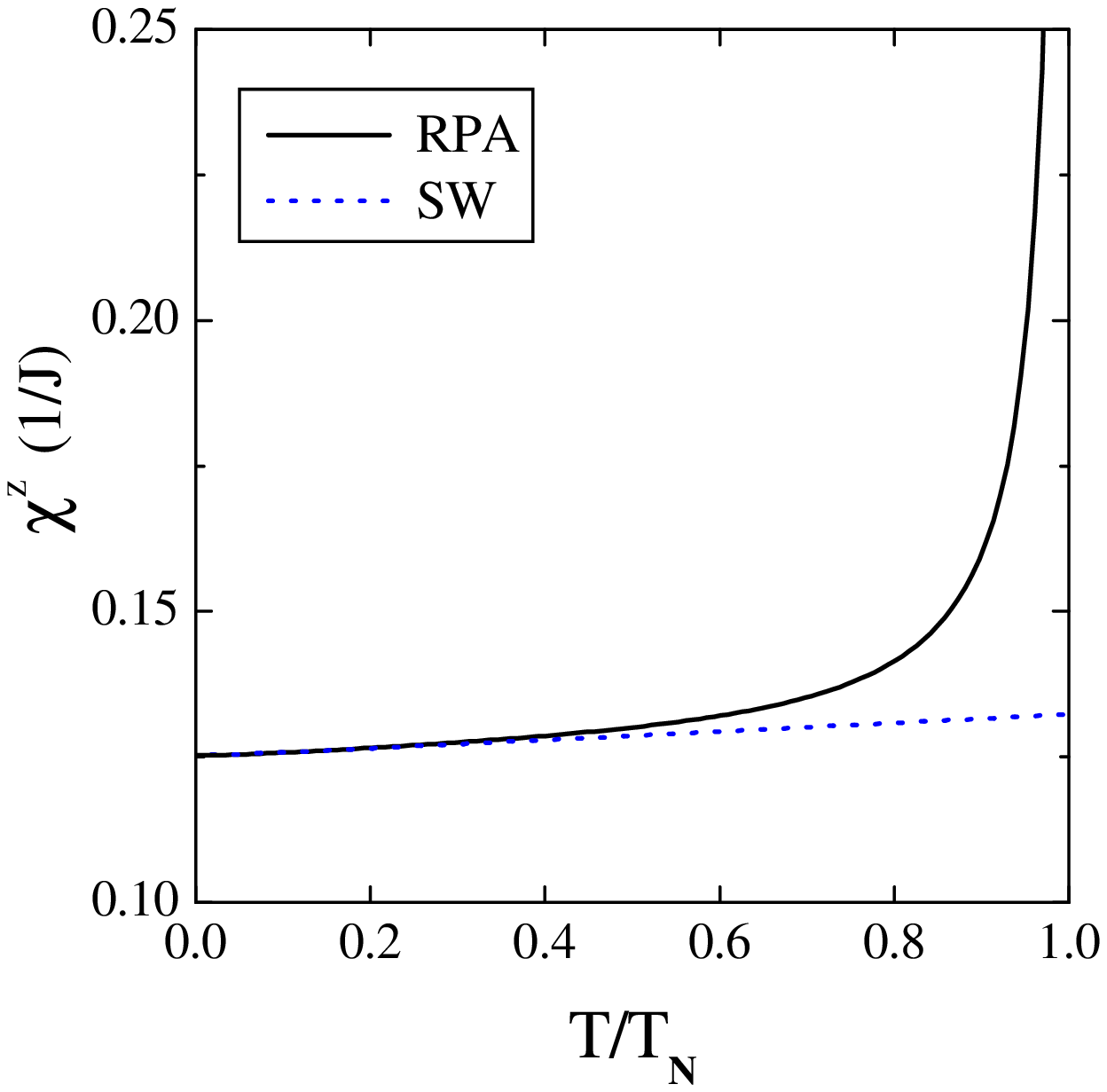}\;(b)
\caption{\label{fig:susc_z} (Color online) The susceptibility $\chi^z$ within
  (a) the RPA (black solid line) and MFA (red dashed line), as well as
  (b) a comparison of $\chi^z$ below the ordering temperature within the RPA
  (black solid line) and spin-wave (SW)
  (blue dotted line) theory. As in previous plots, we have used $d/J = 0.02$ and
  $\Delta\Gamma/J = 0.42\times 10^{-3}$.}
\end{figure}
We find that at low $T$, as was also found for $\chi^y$, the RPA
is in the the good agreement with spin-wave predictions.
Near the transition temperature, the RPA method leads to the qualitatively
different behaviour of the $z$ component of susceptibility with respect to
both the MFA and spin-wave formalisms.

The differences between the MFA \textit{vs.} RPA data shown above can be understood
using the following reasoning. Firstly, consider below the N\'eel temperature. The
canted moments which develop are confined to lie in the $y-z$ plane; as well, they
are ferromagnetically ordered in the $z$-direction (recall that we are studying a
single CuO$_2$ plane). Then, one can see that within the MFA the weak FM produces
a divergence of the $z$-component of the susceptibility only in a very narrow
temperature region close to the N\'eel point.
Since the MFA does not account for near-neighbour correlations between
the spins, away from the immediate vicinity of $T_N$ the weak FM is ignored
and $\chi^z$ behaves like a $T$-independent transverse susceptibility (that
is, transverse to the ordered AF moment).
In contrast to this, the $z$-component of the susceptibility calculated within
the RPA has a strong temperature dependence and shows that the effects of the
quantum fluctuations are important in a wide region below the N\'eel temperature.
The other component which shows some differences between the MFA and RPA
below the transition is $\chi^y$, and for this component it is seen that since the
MFA does not include the reduction of the staggered moment (which is in this
direction) due to quantum fluctuations at low temperatures, linear spin-wave theory,
and not the MFA, agrees with the RPA for low temperatures.

Further, above the N\'eel temperature the differences can be understood
as follows. In a MFA (that is $T_N^{MFA}\approx J$),
both components of the susceptibility $\chi^y$ and $\chi^z$
are rapidly changing functions in the immediate vicinity of $T_N$
($\delta T/T_N \sim 0.005$), and then have the same behaviour as the $\chi^x$ term
further above the transition. This MFA behaviour is in no way similar to
that found in the RPA. That is, our results are an example of the
pronounced effects of short-range correlations and quantum fluctuations.
The RPA scheme gives a much lower value of the N\'eel temperature than MFA does
($T_{\rm N} \sim 0.3J$), but in a broad $T$ region above the N\'eel point
strong short-range correlations exist, and the RPA includes the manner
in which these fluctuations strongly modify the susceptibility. Similar
reasoning explains the differences in $\chi^x$ between the MFA and RPA.

\begin{figure}[h]
 \epsfxsize 0.34\textwidth\epsfbox{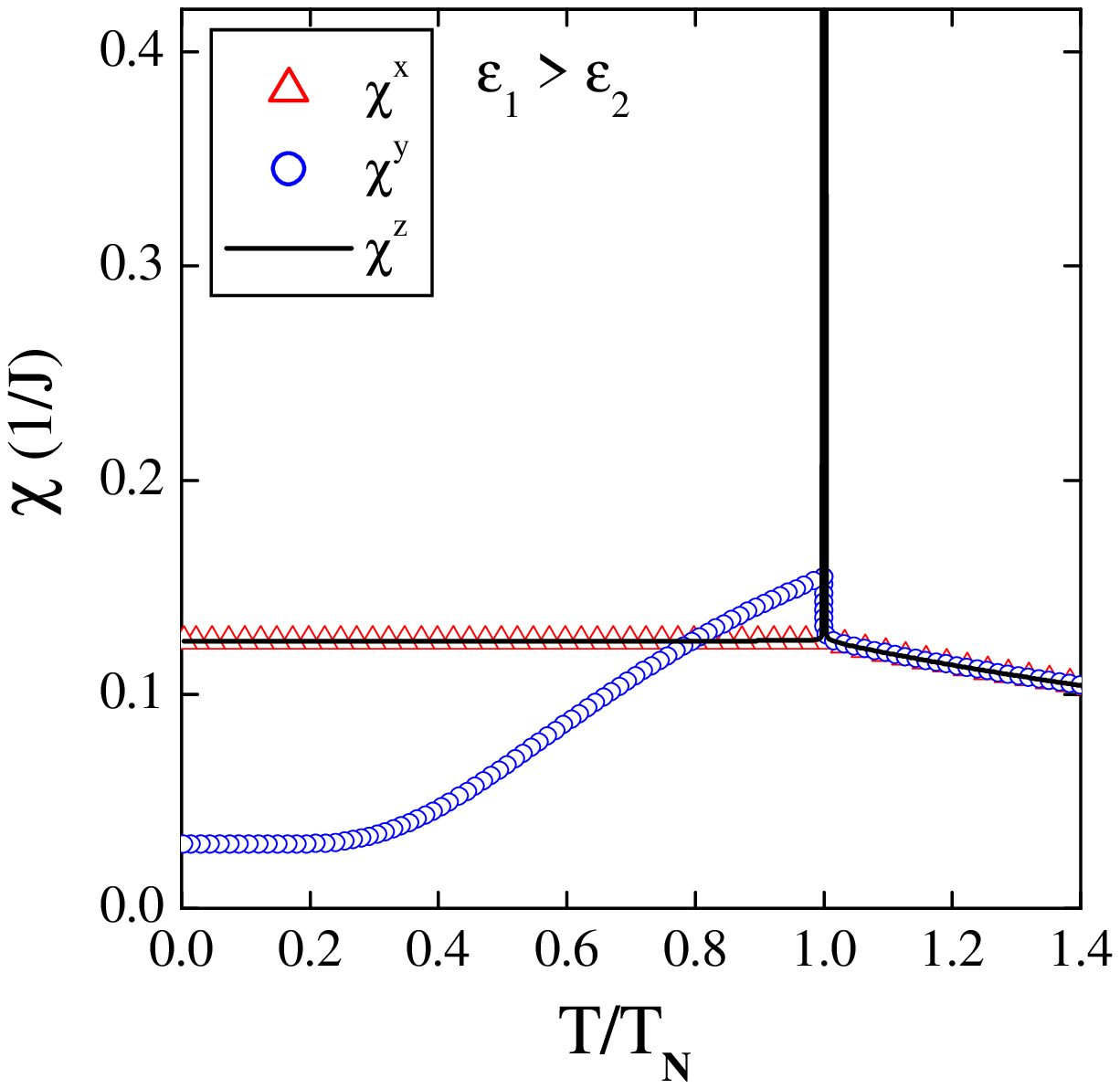}\;(a)\\
 \epsfxsize 0.34\textwidth\epsfbox{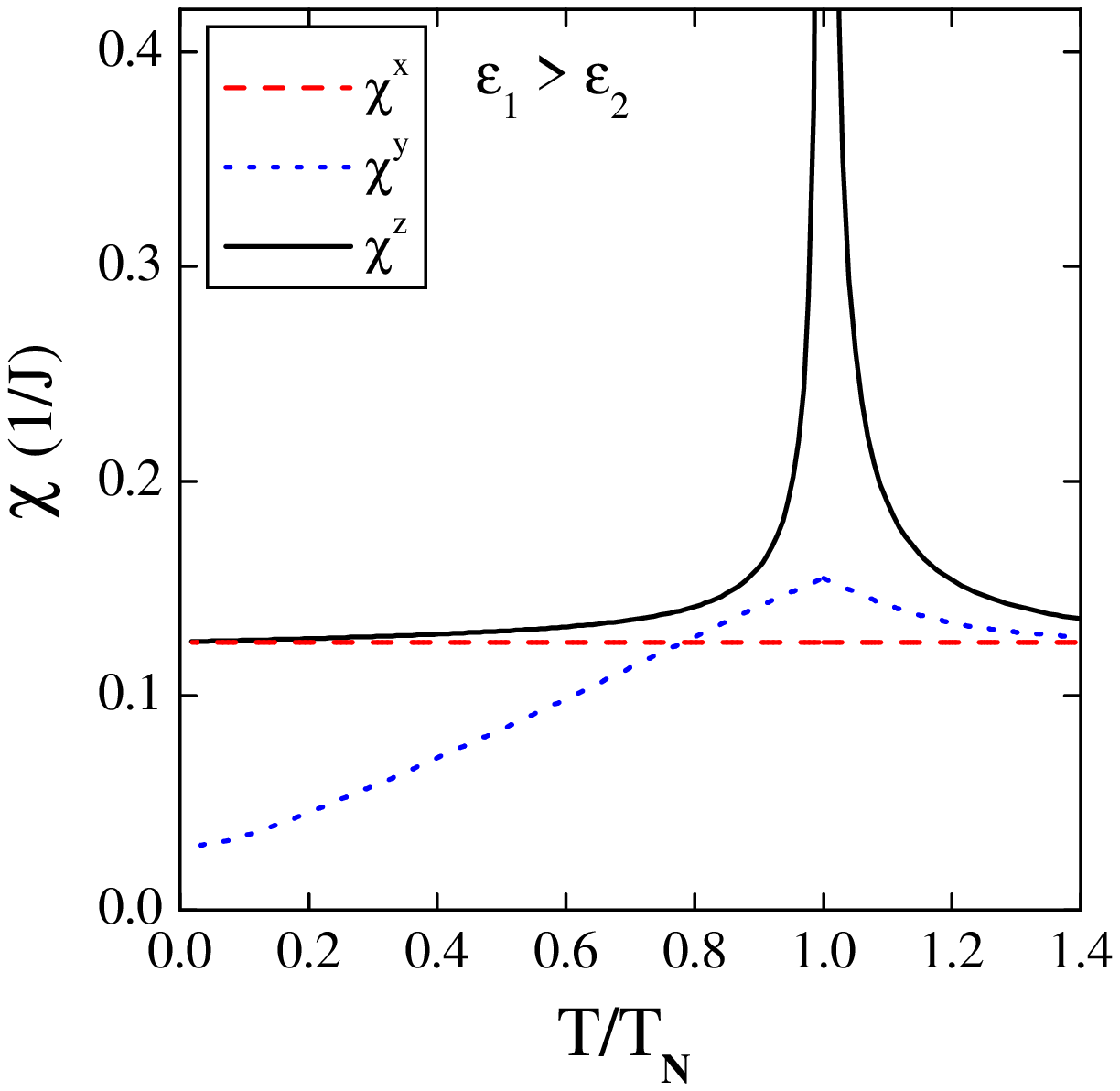}\;(b)
\caption{\label{fig:chis1_MFAnRPA} (Color online)
      All 3 components of the susceptibility within (a) the MFA, and (b) within the RPA,
      for $d/J = 0.02$ and $\Delta\Gamma/J = 0.42\times 10^{-3}$.}
\end{figure}
\begin{figure}[h]
 \epsfxsize 0.34\textwidth\epsfbox{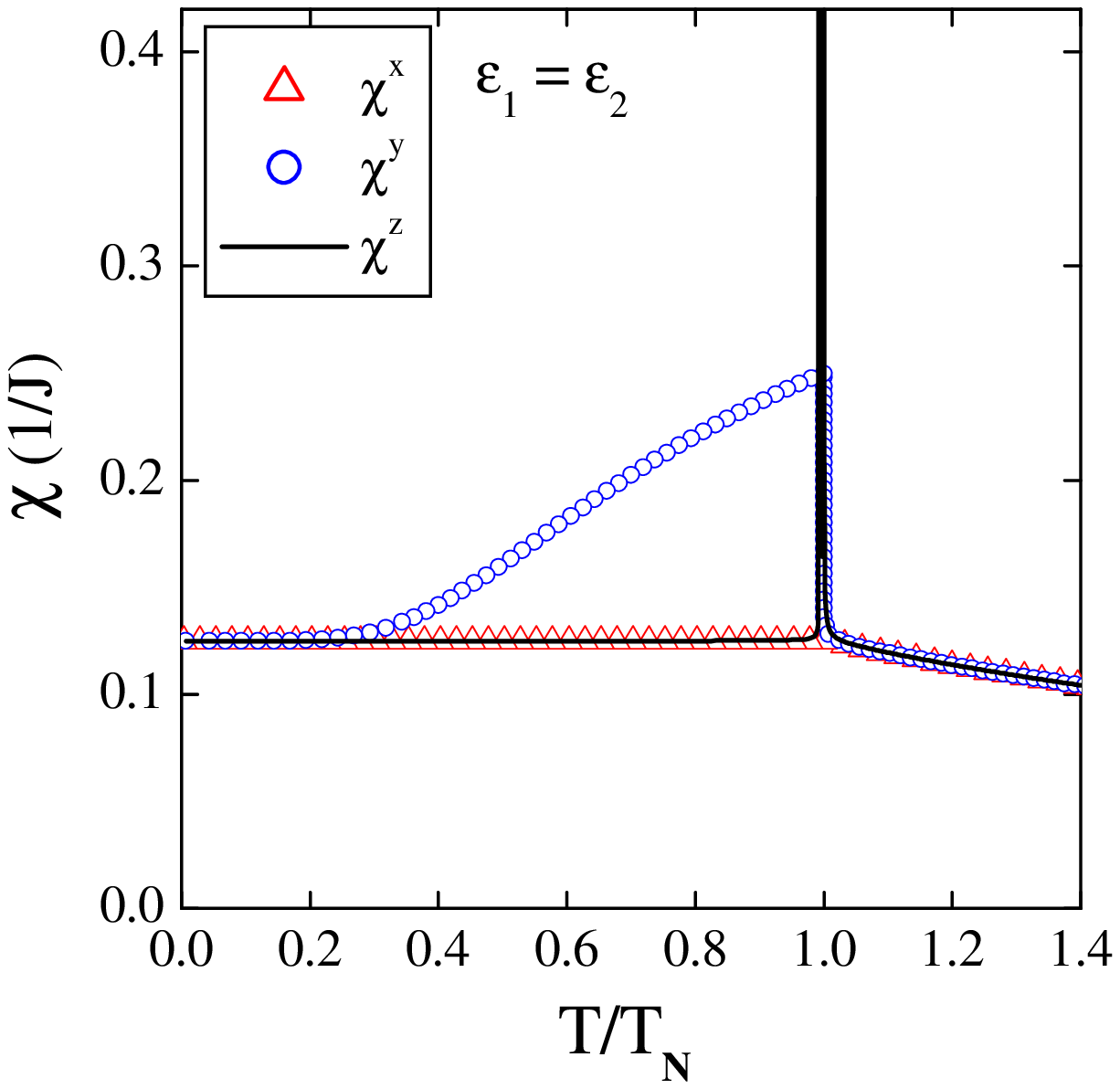}\;(a)\\
 \epsfxsize 0.34\textwidth\epsfbox{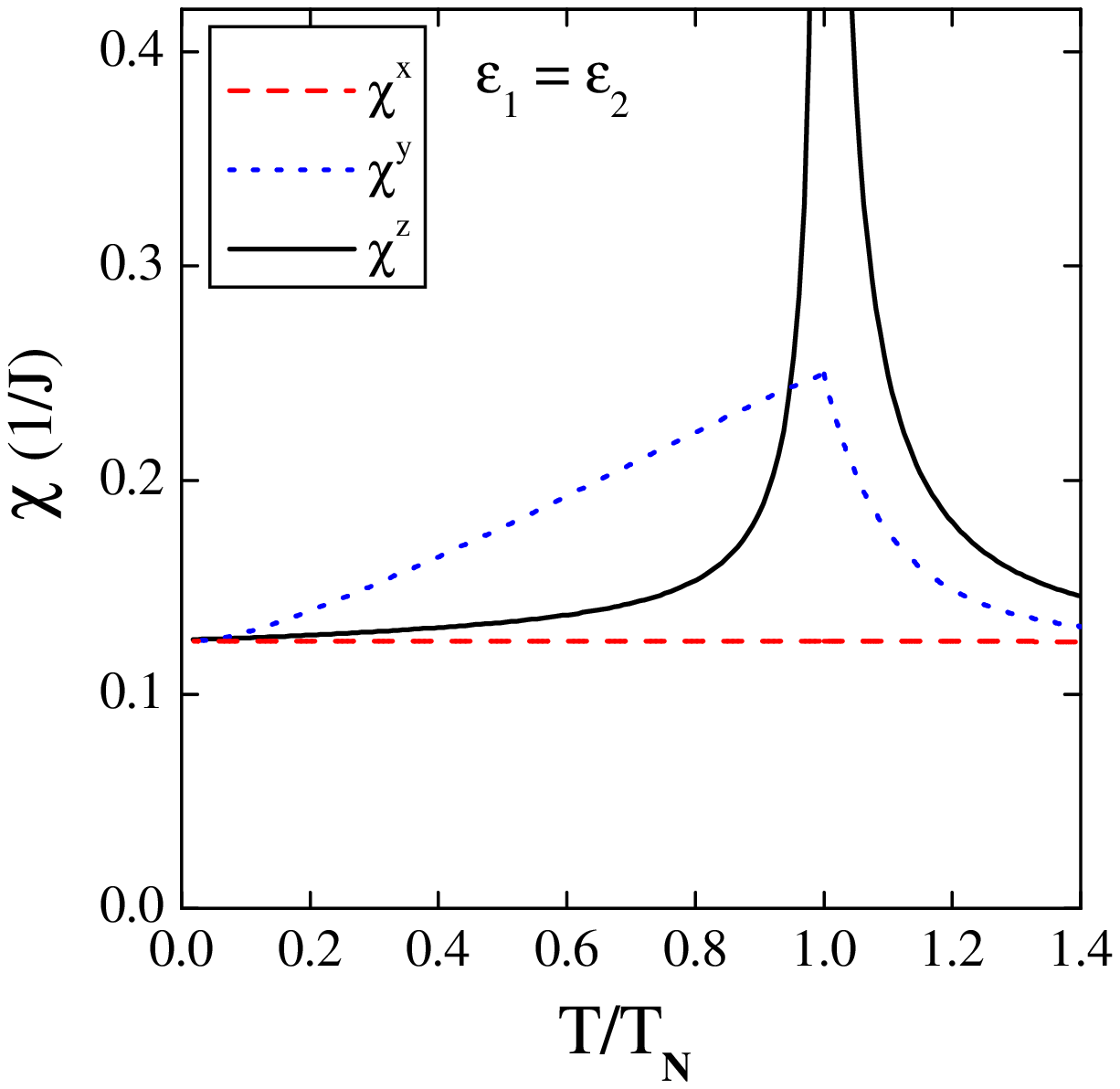}\;(b)
\caption{\label{fig:chis2_MFAnRPA} (Color online)
      All 3 components of the susceptibility within (a) the MFA, and (b) within the RPA,
      for $d/J = 0.041$ and $\Delta\Gamma/J = 0.42\times 10^{-3}$.}
\end{figure}
\begin{figure}[h]
 \epsfxsize 0.34\textwidth\epsfbox{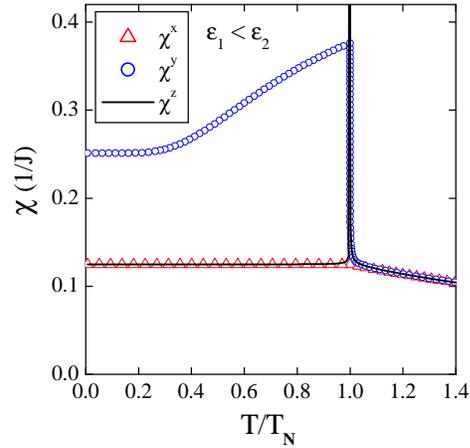}\;(a)\\
 \epsfxsize 0.34\textwidth\epsfbox{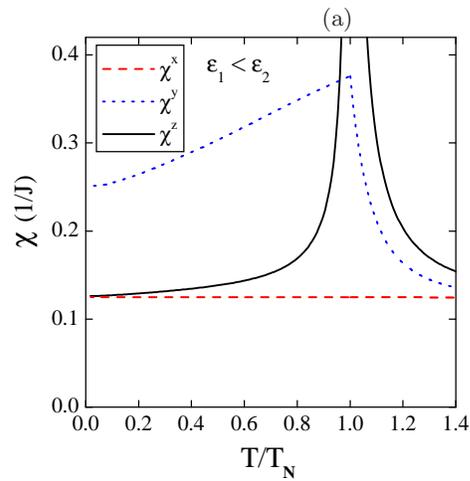}\;(b)
\caption{\label{fig:chis3_MFAnRPA} (Color online)
      All 3 components of the susceptibility within (a) the MFA, and (b) within the RPA,
      for $d/J = 0.058$ and $\Delta\Gamma/J = 0.42\times 10^{-3}$.}
\end{figure}

For completeness, in Figs.~\ref{fig:chis1_MFAnRPA},\ref{fig:chis2_MFAnRPA},\ref{fig:chis3_MFAnRPA}
we present all components of the susceptibility together, within both the MFA and the RPA,
contrasting  different values of the physical parameters describing the DM interaction.
To be specific, in Fig.~\ref{fig:chis1_MFAnRPA} we show the situation when $\varepsilon_1 > \varepsilon_2$
with the ratio $(\varepsilon_1/\varepsilon_2)^2\approx 4.2$ at zero temperature.
As a result, we obtain that for $T=0$ $\chi^y < \chi^x,\chi^z$ with the same ratio between
the $x,z$ and $y$ components of susceptibility $\chi^{x,z}/\chi^y\approx 4.2 $ (see \S\ref{subsec:Teq0chis}).
By increasing the magnitude of the DM parameter $d$, due to the strong dependence of the mode
$\varepsilon_2$ of the $d$ (see Fig.~\ref{fig:gaps_d}), we obtain the situation corresponding to
$\varepsilon_2 = \varepsilon_1$.
Then, as is seen in Fig.~\ref{fig:chis2_MFAnRPA}, both the MFA and RPA schemes result
in equal values of the all components of susceptibility at low $T$.
A further increasing of$d$ leads to the situation $\varepsilon_2 > \varepsilon_1$,
opposite to the one presented in Fig.~\ref{fig:chis1_MFAnRPA}.
In Fig.~\ref{fig:chis3_MFAnRPA} we show the susceptibility in the case of the ratio
$(\varepsilon_2/\varepsilon_1)^2\approx 2.0$, and  at $T=0$ one finds
$\chi^y > \chi^x,\chi^z$ and $\chi^y/\chi^{x,z}\approx 2.0$ for $T=0$.
(We note that for other sets of $d,\Delta\Gamma$, the behaviour of the components
of $\chi$ is determined almost entirely by the ratio of the spin-wave gaps, $\varepsilon_1/\varepsilon_2 = $.
These results agree with our analytical predictions (see section~IV).)

Then, comparing the $z$-component of the susceptibility within the
RPA (Figs.~\ref{fig:chis1_MFAnRPA}(b)-\ref{fig:chis3_MFAnRPA}(b))
we also find that increasing of the anisotropy parameter $d$ leads to the broadening $T$ regions
where (i) the effects of the quantum fluctuations are important $T<T_N$ and
     (ii) the strong short-range correlations exists $T>T_N$.

\section{Summary and conclusions/discussion}
\label{sec:conclusions}

To summarize, we have presented a theoretical investigation of a single \CuO~plane
of the undoped \LCO~crystal in the low-$T$ orthorhombic phase.
The Cu spins in the plane were modelled by the 2D spin-1/2 Heisenberg AF with spin-orbit
coupling, the latter represented the antisymmetric and symmetric DM anisotropies.
We have adopted the Green's function method within Tyablikov's RPA decoupling scheme to calculate
the magnetic susceptibility of such a model. In order to allow us to accurately model
the longitudinal susceptibility within such a level of decoupling of high-order
Green's functions, we have extended Liu's method \cite{Liu} for the isotropic Heisenberg
model to one that includes a weak canted FM moment in the plane.

We can emphasize several important conclusions from our results. We have found that the
anisotropy introduced into the problem by the symmetric and anti-symmetric DM interactions
leads to important changes in the behaviour of the magnetic susceptibility near the transition point.
By comparing the MFA and RPA results we conclude that the effects of quantum fluctuations and
the short-range correlations are very strong in the {\em {broad temperature region}} of near the
N\'eel temperature. Further, we find that since the RPA and SW results are quite different near
the N\'eel temperature, the effects of spin-wave interactions, which are included in an
approximate way in the RPA but not the SW theories, are very important in this system.
This necessarily leads to the question, would more advanced decoupling schemes, namely
improvements on Tyablikov's decoupling (\textit{e.g.},~see our Eq.~(\ref{eq:Tyablikov_decoupling})),
or, possibly, the inclusion of nonlinear effects in the SW theory, lead to qualitatively different results?

Secondly, we have obtained that the weak ferromagnetism in the $z$-direction (caused by the DM interaction)
leads to the essential difference between the temperature behaviours of the transverse $\chi^x$ and $\chi^z$
components of the susceptibility
(recall that the AF moments lie in the $y-z$ plane and are nearly aligned along the $y$ axis).
We established the correlation between the ratio of the in- and out-of-plane spin-wave
modes of the excitation spectrum in the long wavelength limit (${\bk}=0$),
which is fixed by the ratio between the $d$ and $\Gamma_1-\Gamma_3$ DM parameters,
and the behaviour of $\chi^{x,z}$ \textit {vs.}~$\chi^y$ in the zero temperature limit.
This conclusion is independent on the analytical method which we used to calculate
the susceptibilities, since all methods agree in the low-$T$ regime, and could
allow one to make predictions concerning the gaps in the excitation spectrum
based on the data for the susceptibility.

Now we comment on the comparison of our results to the experimentally observed anisotropies \cite {Ando}
that motivated this work. We can state that, in addition to the known results\cite{Coffey2,Aharony1,Koshibae}
that DM interaction induces the weak ferromagnetism in the LTP phase and the spin-wave gaps, this interaction
is at least in part responsible for the unusual anisotropy in the magnetic susceptibility.\cite{Ando}
We can mention the most significant features observed in the experiment that are in qualitative agreement
with the presented in paper theoretical results: (i) the absence of any special behaviour (anomaly) in the
transverse component $\chi^x$ across the N\'eel temperature; (ii) the additional increase of the $\chi^y$
component in the ordered state and its smooth decrease in a broad temperature region in the paramagnetic
state; (iii) a significant temperature dependence of the component $\chi^z$ in the broad temperature
region below and above the transition point.

Now we briefly discuss the experimental data which cannot be explained within the framework of the
proposed here theory. Firstly, we have found that the observed ratio between the $x$ and $y$ components
$\chi^x<\chi^y$ (in the $T=0$ limit) takes place only if the spin-wave gap with out-of-plane
mode is less than the in-plane one $\varepsilon_{\rm o}<\varepsilon_{\rm i}$. However, older neutron-scattering
experiments\cite{Keimer} find the opposite ratio: the magnitude for the out-of-plane mode is 5~meV, for the
out-of-plane mode 2.3~meV. Recent Raman work confirms one of these values.\cite{gozar04}
So, other interactions which affect these gaps must be important for an accurate explanation of the
susceptibility data. Secondly, our results cannot explain a $T$-independent shift between $\chi^x,\chi^y$ and
$\chi^z$ observed in experiments -- an explanation of this physics is provided in the experimental
paper, namely that one must include a van Vleck contribution which shifts, in a $T$-independent manner,
these components of the susceptibility, but we defer our inclusion of this physics until the second
paper in this series of theoretical studies.

For further improvements of our theoretical modelling of the \LCO~compound, it seems to be important
to investigate a 3D model on a body-centered lattice with the weak AF interlayer coupling.
It is also possible to extend the 2D model by considering the ring exchange, and the interaction between
the next nearest neighbour sites, and
we expect that some of these additional physics can be responsible for the correct ratio between the
spin-wave gaps with respect to the ratio between $\chi^x$ and $\chi^y$.
In addition, the anisotropic Van Vleck contribution (orbital susceptibility) and gyromagnetic (Land\'e)
factor need to be taken into account.
We will present a detailed comparison to these experiments when these other interactions are included
in future publications.

\begin{acknowledgments}

We wish to thank Yoichi Ando, Alexander Lavrov and David Johnston for helpful discussions.
One of us (RJG) thanks the Aspen Center for Physics, where part of this work was completed.
This work was partially supported by the NSERC of Canada and NATO.

\end{acknowledgments}

\bibliographystyle{unsrt}
\bibliography{LTPaper_Jan0405}

\end{document}